\documentclass[11pt]{article}

\def\diag{{\rm diag}}
\def\pconv{\smash{\mathop{\longrightarrow}\limits^p}}     
\def\dconv{\smash{\mathop{\longrightarrow}\limits^d}}     

\def\diag{\mbox{diag}}

\renewcommand{\bar}{\overline}

\newcommand{\To}{\longrightarrow}

\renewcommand{\tilde}{\widetilde}
\renewcommand{\hat}{\widehat}

\def\plim{\mbox{plim }}

\newcommand{\beq}{\begin{eqnarray*}}
\newcommand{\eeq}{\end{eqnarray*}}

\def\1T{frac{1}{T}}
\def\1n{\frac{1}{n}}

\def\bitem{\medskip\begin{itemize} \itemsep=8.0pt \parskip=8.0pt}
\def\eitem{\end{itemize}}

\usepackage{fullpage,amssymb,amsmath,amsthm,mathtools,natbib,soul,xcolor,url,enumitem,calc}
\newcommand\x{\checkmark}
\newcommand\y{\times}
\definecolor{Gray}{gray}{0.9}
\usepackage{tikz}
\usetikzlibrary{matrix,fit}
\usetikzlibrary{backgrounds}

\newtheorem{lemma}{Lemma}
\newtheorem{proposition}{Proposition}

\def\pr{{^\prime }}
\def\tall{\textsc{tall}}

\def\pconv{\smash{\mathop{\longrightarrow}\limits^p}}     
\def\dconv{\smash{\mathop{\longrightarrow}\limits^d}}     
\def\pconv{\smash{\mathop{\longrightarrow}\limits^p}}     
\def\dconv{\smash{\mathop{\longrightarrow}\limits^d}}     
\newcommand{\No}{N_o}

\newcommand{\oi}{{o_i}}

\begin{document}

\title{Factor-Based Imputation of Missing Values and Covariances in Panel Data of Large Dimensions}

\author{Ercument Cahan\thanks{Independent researcher,  Email: ecahan@gmail.com}
\and
Jushan Bai\thanks{Department of Economics, Columbia University, 420 W. 118 St. MC 3308, New York, NY 10027. Email: jb3064@columbia.edu}
\and
Serena Ng\thanks{Department of Economics  Columbia University and NBER, 420 W. 118 St. MC 3308, New York, NY 10027. Email: serena.ng@columbia.edu. \newline
We thank Bennie Chen for excellent research assistance.
This paper was presented at  2021 SOFIE webinar, the 2021 Econometric Society Asia meeting, and the 2021 Statistics Canada meeting, and the econometrics and statistics seminar at Chicago Booth. We thank seminar participants, Markus Pelger, Andrew Patton,  Dacheng Xiu,  two anonymous referees, the associate editor and co-editor (Torben Andersen) for helpful comments.    Ng acknowledges financial support from the National Science
Foundation  SES-1558623 and SES-2018369.}
}
\date{ \today }

\maketitle

\begin{abstract}
Economists are blessed with a wealth of data for analysis, but  more often than not,  values in some entries of the data matrix are missing.  Various methods have  been proposed  to handle   missing observations in a few variables.  We  exploit the factor structure in  panel  data of large dimensions. Our \textsc{tall-project} algorithm first estimates the factors from a \textsc{tall} block in which data for all rows are observed, and  projections of unit specific length are then used to estimate the factor loadings. A missing value is imputed by  its  estimated  common component which we show is consistent and asymptotically normal   without further iteration. Implications for using imputed data in factor augmented regressions are then discussed.
 To compensate for the downward bias in sample covariance matrices created by an omitted noise in each imputed value, we    overlay the imputed data with re-sampled idiosyncratic residuals many times and use the average of the  covariances to estimate the parameters of interest. Simulations show that the procedures have desirable finite sample properties.

\end{abstract}
\noindent Keywords: risk management, covariance structure,  matrix completion, incomplete data

\noindent JEL Classification: C1, C2.

\bibliographystyle{harvard}
\setcounter{page}{0}
\thispagestyle{empty}

\newpage
\baselineskip=18.0pt

\section{Introduction}
Missing data is a problem that empirical researchers frequently encounter. For example, data can be missing due to  attrition in longitudinal surveys such as  the PSID and SIPP,  to  IPOs and  bankruptcies in the case of stock returns, and to staggered releases of data by different institutions. Data sampled  at a lower  frequency will have missing values in an analysis that involves higher frequency data. Whatever is the reason, the way we handle missing data is not innocuous. Dropping rows or columns necessarily entails loss of information, and while the EM algorithm  allows  imputation of missing values, it is designed for low dimensional data and may not be easily scalable.

In recent years, progress has been made for situations when the  missing values occur  in large panels of data with a low rank component.
Assuming incoherence conditions  to ensure that the low rank structure is non-degenerate  as in \citet{candes-recht:09}, the machine learning literature has developed  regularization based  algorithms  to solve  matrix completion  problems of the  Netflix challenge type. On the econometrics front, recent work by \citet{su-missing}  \citet{xiong-pelger:19}, and \citet{baing-miss:jasa} propose different implementations  that permit the entire common (low rank) component to be consistently estimated  if the  factor structure is strong.  While the machine learning literature provides worst-case error bounds, the econometrics literature provides the distribution theory that is needed for inference, and is also the approach that this paper takes.

We introduce a \text{Tall-Project} estimator  (or \text{TP} for short) for imputing missing values in a panel of data with $T$ rows and $N$ columns, where $N$ and $T$ are both large. The factors are estimated from  a \tall\ block consisting of  complete data for $N_o\le N$ units, while  the loadings  are obtained from  $N$ time series  projections  with series-specific sample size. We obtain two main results. First, under certain  assumptions on the factor structure in the different blocks,   we show  that the  \textsc{tp} estimates  are  consistent and asymptotically normal for every entry of the low rank component  though the convergence rate is series specific. And while iteration is not needed for consistent estimation, one re-estimation can improve the convergence rate. Because of missing data, stronger conditions are needed for the factor estimates to be treated as known in  factor-augmented  regressions. \footnote{An \textsc{R} package that implements \textsc{tw} and \textsc{tp} is available for download at \url{htpps://github.com/cykbennie/fbi.}}

Our second result concerns estimation of covariances  from  imputed data.  Covariances play an important role  in portfolio analysis and  in structural equation  (covariance structure) modeling.   While  the imputed data  is unbiased for its mean,  the variance of the imputed data  is  biased because the idiosyncratic noise associated with the missing observations are  set to zero.  To remedy this problem,   we  repeatedly  overlay the first step estimate of missing values  with  resampled   idiosyncractic residuals before computing sample covariances. An  average of these imputed covariances is then used to estimate the objects of interest. In  simulations calibrated to  CRSP data from 1990-2018,  resampling from the own residuals (ie. without  pooling)  yields risk measures that compare favorably and sometimes outperform the ones based on a factor-based covariance estimator.

In what follows, let $X$ be a $T\times N$ panel of data,   $X_i=(X_{i1},\ldots X_{iT})^\prime$ be a $T\times 1$ vector of random variables and
 $X=(X_1,X_2,\ldots, X_N)$ be a $T\times N$  matrix.
We use  $i=1,\ldots N$ to index cross-section units and $t=1,\ldots T$ to index time series observations.  In practice, $X_{i}$ is transformed to be stationary, demeaned, and is often standardized. It is assumed that  the normalized data $Z=\frac{X}{\sqrt{NT}}$
has  singular value decomposition (\textsc{svd})
\[ Z=\frac{X}{\sqrt{NT}}
 =U DV\pr=U_r D_r V_r\pr+ U_{n-r}  D_{n-r} V_{n-r}\pr\]
where $ D_r$ is a diagonal matrix of $r$ singular values  ordered such that $ d_1\ge d_2\ldots \ge d_r$, while $U_r, V_r^\prime$ are the corresponding left and right singular vectors respectively. Analogously, $U_{n-r}$ and $V_{n-r}^\prime$ are $n\times (n-r)$  matrices of left and right eigenvectors associated with $d_{r+1},\ldots, d_n$. The low rank component  $U_rD_rV_r^\prime$ can be defined without probabilistic assumptions..

We consider a factor model defined as
    \begin{eqnarray}
    \label{eq:dgp}
    X&=&F\Lambda^{\prime} + e
    \end{eqnarray}
 where  $F$ is a $T\times r$ matrix of common factors,  $\Lambda$ is a $N \times r$ matrix of factor loadings, and $e$ is a $T\times N$ matrix of  idiosyncratic errors.  We will let  $(F^0,\Lambda^0)$ be the true values of $(F,\Lambda)$. The common component $C^0=F^0\Lambda^{0\pr}$ has reduced rank $r$ because $F^0$ and $\Lambda^0$ both have rank $r$. We will estimate the  factors  and loadings   by the method of static asymptotic principal components (APC).\footnote{In this paper, we rely heavily on the  results  given in \citet{stock-watson-jasa:02,baing-ecta:02,bai-ecta-03,baing-ecta:06}.}
The normalization $\frac{F\pr  F}T=I_r$ gives the APC estimates that are   unique  up to a column sign:
\[ (\tilde F,\tilde\Lambda)=(\sqrt{T} U_r, \sqrt{N} V_r D_r). \]
 Since $r$ can be consistently estimated, we proceed as though $r$ is known.

We consider a large dimensional approximate factor model in which $T$ and $N$ are large and which satisfies the following assumptions:

\paragraph{Assumption A:} There exists a constant $M<\infty$ not depending on $N,T$ such that
\begin{itemize}
    \item[a.] (Factors and Loadings):
    \begin{itemize}
      \item[(i)]$E\|F_t^0\|^4 \le M$, $\|\Lambda_i\|\le M$;
      \item[(ii)]
$\frac{F^{0\pr}F^0}{T}\pconv \Sigma_F >0$, and $ \frac{\Lambda^{0\pr}\Lambda^0 }{N}\pconv \Sigma_{\Lambda}>0$;
\item[(iii)] the eigenvalues of $\Sigma_F \Sigma_{\Lambda}$ are distinct.
\end{itemize}
\item[b.] (Idiosyncratic Errors): Time and cross-section dependence
\begin{itemize}
\item[(i)] $E(e_{it})=0, E|e_{it}|^8\le M$;
\item[(ii)] $E(\frac{1}{N}\sum_{i=1}^N e_{it} e_{is})=\gamma_N(s,t)$, $\sum_{t=1}^T |\gamma_N(s,t)|\le M$,  $\forall s$; 
\item[(iii)] $E(e_{it}e_{jt})=\tau_{ij,t}$, $|\tau_{ij,t}|\le | \tau_{ij}|$ for some $ \tau_{ij}$  $\forall t$, and
$\sum_{j=1}^N |\tau_{ij}|\le M$, $\forall i$;
\item[(iv)] $E(e_{it}e_{js})=\tau_{ij,st}$ and $\frac{1}{NT}\sum_{i=1}^N \sum_{j=1}^N \sum_{t=1}^T \sum_{s=1}^T|\tau_{ij,ts}|<M$;
\item[(v)]  $E|N^{-1/2}\sum_{i=1}^N  [e_{is}e_{it}-E(e_{is}e_{it})]^4\le M$ for every $(t,s)$;
\item[(vi)] $E(\|\frac{1}{\sqrt{T}} \sum_{t=1}^T F_t^0e_{it}\|^2)\le M, \forall i$, and $E( \|\frac{1}{\sqrt{NT}} \sum_{i=1}^N \sum_{t=1}^T \Lambda_i^0 F_t^{0\prime} e_{it}\|^2)\le M$.
\end{itemize}
\item[c.] (Central Limit Theorems): for each $i$ and $t$,
$   \frac{1}{\sqrt{N}} \sum_{i=1}^N \Lambda^0 _i e_{it}\dconv N(0,\Gamma_t)$ as  $N\rightarrow\infty$, and
$   \frac{1}{\sqrt{T}}\sum_{t=1}^TF^0_te_{it}\dconv N(0,\Phi_i) $ as  $T\rightarrow \infty$; the two limiting distributions are independent.

\end{itemize}

Assumption A is used in  \citet{bai-ecta-03} to develop an inferential theory for large dimensional factor analysis when $X$ is completely observed.  The  moment conditions  in (b) ensure  that the factor structure is strong and can be separated from the idiosyncratic errors which can be weakly correlated, both in the time and the cross-section dimensions. Let $\mathbb D_{r}^2$ and $ \mathbb V_{r } $ be the eigenvalues and eigenvectors of the $r\times r$ matrix $\Sigma_{\Lambda}^{1/2}\Sigma_F\Sigma_{\Lambda}^{1/2}$, respectively.   As shown in \citet{bai-ecta-03},  $\plim_{N,T\rightarrow\infty} D_r^2 = \mathbb D_r^2$ and $\plim_{N,T\rightarrow\infty} \frac{\tilde F\pr F^0}{T}=\mathbb Q_r$,
 where
  $\mathbb Q_r=\mathbb D_{r} \mathbb V_{r}\Sigma_{\Lambda}^{-1/2}$. The following  properties of the APC estimator are given in  \citet{bai-ecta-03}:

    \begin{lemma}
      \label{lem:bai-03}
 Suppose that Assumption A holds and let $H=(\Lambda^{0'}\Lambda^0/N)(F^{0'}\tilde F/T) D_{r}^{-2}$.
If $\sqrt{N}/T\rightarrow 0$ as $ N,T\rightarrow \infty$, then
\begin{subequations}
\begin{eqnarray}
\sqrt{N} (\tilde F_t-  H\pr  F_t^0)&
\dconv&  \mathcal N\bigg(0,   \mathbb D_{r}^{-2}\mathbb Q_r \Gamma_t \mathbb Q_r\pr \mathbb D_{r}^{-2}\bigg)\label{eq:F-full}\\
\sqrt{T} (\tilde \Lambda_i-  H^{-1} \Lambda_i^0)&
\dconv & \mathcal N\bigg(0,  (\mathbb Q_r\pr)^{-1} \Phi_i \mathbb Q_r^{-1} \bigg)\label{eq:Lambda-full}\\
  \min(\sqrt{N},\sqrt{T})\bigg(\frac{\tilde C_{it}-C_{it}^0 }{\sqrt{\tilde{\mathbb V}_{\tilde C_{it}}}}\bigg) &\dconv& \mathcal N(0,1). \label{eq:C-full}
\end{eqnarray}
\end{subequations}
where
$    \tilde {\mathbb V}_{\tilde C_{it}}( N, T)$ is a consistent estimate of $ \frac{\delta^2_{ N T}}{ N}\Lambda_i^{0\prime} \Sigma_{  \Lambda}^{-1}  \Gamma_t \Sigma_{\Lambda}^{-1}\Lambda^0_i+
     \frac{\delta^2_{ N T}}{ T}F_t^{0\prime} \Sigma_{F}^{-1}  \Phi_i \Sigma_{F}^{-1}F^0_t $.
\end{lemma}
In what follows, we will obtain results for the factor estimates when $X$ has missing values.

\section{Missing Data}
 A panel is said to be  \textit{complete} if it does not have missing values. In practice most data panels have missing values. The data are said to be {\em missing completely at random} (MCAR) if missingness is  independent of the data whether or not they are observed.  As  the observed and missing  data  are  drawn from the same underlying  distribution and hence have no systematic differences, MCAR does not create bias.  This is not the case  with   {\em missing at random} (MAR), though  the systematic  differences  can  be explained by other observables. For example,   missing values in financial statements of smaller firms may arise   because smaller firms are less regulated, in which case  the missing data propensity   can  be related to   market capitalization  which is observed for publicly traded firms. But the issue remains that  the observed values under  MAR  do not form  a random sample.

To overcome the bias when the data are missing at random,  one approach  is to re-weigh the data (such as  'inverse probability weighting' used in survey design).  A second approach is to rebalance  the panel  by amputation which can take one of three forms: (a) listwise (complete case)  deletion under  which  the entire row with missing values is eliminated;
(b) variable deletion under which a series with one or more  missing values will be eliminated;
and (c) pairwise deletion where  only the cases  with missing data involved will be deleted. Though  the solution is simple and does not entail modeling assumptions, amputation leads to significant information loss.
Furthermore,  the  resulting  balanced  sample  may not be representative of the population  and may yield biased estimates.

Another way to rebalance  the panel is   imputation.
The simplest  procedure is to replace the missing values with the sample mean or median. A more sophisticated procedure is to impute from   the unconditional sample first and second moments via the EM algorithm. \citet{schneider:01} proposes a regularized EM algorithm that is quite popular in analysis of climate data.  Imputation on the basis of unconditional moments  does not require a model. A fully specified approach is to  construct the likelihood using incomplete observed data and iteratively solve for the parameters of the conditional mean function.   The estimates would be more efficient if the parametric assumptions were correct.
These methods are designed for imputing missing values in  a small number of variables/predictors. See \citet{robins-rotnitzky-zhao:95} \citet{li-shen-li-robins:13} and  \citet{raghunathan:04} for a review.

 Factor models provide a simple framework for imputing missing values of many variables.   \citet{banbura-modugno:14}, \citet{jungbacker-koopman}, \citet{jkv:11} consider state-space modeling   of a strict factor (i.e., a diagonal idiosyncratic error variance matrix) with missing data and uses  non-linear filters to compute the likelihood as $F_t$ and $\Lambda_i$ are both random.
\citet{giannone-reichlin-small} initializes the missing values with   estimates from the balanced panel and  uses the Kalman filter to perform updating. \citet{stock-watson:handbook-16} discusses the issues with state space estimation of factor models with missing data.
  For large dimensional approximate factor models,  \citet{stock-watson-di-wp} suggests  to fill missing values in $X$  with the  APC estimates of the common component. These are all EM algorithms that use  the factor structure to evaluate the conditional mean in the E-step, and  principal components estimation in the M step.


Though many factor-based imputation methods have been used for some time,  the theoretical  properties of the imputed estimates are studied only recently in   independent works  by \citet{su-missing}, \citet{xiong-pelger:19}, and \citet{baing-miss:jasa}.   All three exploit the strong factor structure in a large panel setting  to estimate   the factors and the loadings from incompletely observed  data.  \citet{su-missing} focuses on the case of  missing at random and analyzes the  EM  estimator considered in \citet{stock-watson-di-wp}. After initializing  the missing values to zero,  the factors are  estimated from  data reweighted by the  frequency of no missing values.  Though these  estimates are consistent, they are  not asymptotically normal without further iteration.   \citet{xiong-pelger:19} also   re-weights the  data in  principal components estimation of the loadings, but uses  cross-sectional regressions to  estimate  the factors at every $t$.  Similar to \citet{su-missing},  the estimates are consistent and asymptotically normal when iterated until convergence. but the `all purpose' estimator  proposed in \citet{xiong-pelger:19}   allows for mechanisms other than missing at random. Though robustness  comes  at the cost of larger variances, efficiency can be improved if stronger assumptions on missingness are made.

\begin{figure}[ht]	
\caption{Missing Data with a Well-Organized Structure}
\label{fig:fig1}
\begin{center}
\begin{tabular}{cc}
 Block Missing  &  Semi-Block Missing \\

\begin{tabular}{ c|c| c | c | c | c |  c| }		
      \multicolumn{1}{r}{} &
      \multicolumn{1}{c}{$X_{1}$}
      & \multicolumn{1}{c}{$\ldots$}    &\multicolumn{1}{c}{$X_{N_o}$} &
      \multicolumn{1}{c}{$X_{N_{o+1}}$} &
       \multicolumn{1}{c}{$\ldots,X_{N-1}$} &
        \multicolumn{1}{c}{$X_{N}$} \\ \cline{2-7}
			& &   &    &    &    &         \\  \cline{2-7}
			& &   &    &   &    &          \\  \cline{2-7}
			& &   &    & x   & x    & x             \\ \cline{2-7}
			& &   &    &  x  &  x  &    x         \\ \cline{2-7}
			& &   &    &  x   & x   &    x         \\ \cline{2-7}
	\end{tabular}
		
	& \quad\quad

\begin{tabular}{ c|c |c | c | c | c |  c| }
  \multicolumn{1}{r}{} &
      \multicolumn{1}{c}{$X_{1}$}
      & \multicolumn{1}{c}{$\ldots$}    &\multicolumn{1}{c}{$X_{N_o}$} &
      \multicolumn{1}{c}{$X_{N_{o+1}}$} &
       \multicolumn{1}{c}{$\ldots,X_{N-1}$} &
        \multicolumn{1}{c}{$X_{N}$} \\ \cline{2-7}

			& &   &    &    &    &         \\  \cline{2-7}
			& &   &    &   &    &          \\  \cline{2-7}
			& &   &    &    &     & x             \\ \cline{2-7}
			& &   &    &    & x  &    x         \\ \cline{2-7}
			& &   &    &  x   &    &   x         \\ \cline{2-7}
		\end{tabular}
	
	\end{tabular}
  \end{center}

\end{figure}
While \citet{su-missing} and \citet{xiong-pelger:19}  reweigh the data  in APC estimation,  \citet{baing-miss:jasa} implicitly re-oganizes the   data   into  blocks. This is partly motivated by the concern  that the  missing at random assumption may at times  be inappropriate. Examples of   microeconomic applications are discussed in \citet{abdik}. Furthermore,   for macroeconomic panels such as FRED-MD  where the data are collected  from different sources, assuming that the missing values are due to the same  missing mechanism  seems restrictive.

Consider the left panel of Figure \ref{fig:fig1}.  This systematic (non-random)  missing pattern can arise if,  for example,  weekly data for related variables (such as inventories and orders) are released in the same week of every month. The  right panel shows a case of  `semi-block missing'. This can arise if, for example in a survey, respondent $N$  dropped out early,  respondent $N-1$  did not respond in period $T-1$, while $N_o+1$  did not respond in period $T$.   The observations can be  missing for different reasons likely unknown to the researcher.

While the  missing data mechanism can be difficult to verify, we do observe the missing data pattern, and
organizing the data into blocks offers  a new  way of thinking about imputation.  The insight in \citet{baing-miss:jasa} is that the quantities needed for imputation  can be obtained from two  completely observed blocks.  Their \textsc{tall-wide} (or \textsc{tw} for short)  procedure  uses the \textsc{tall} block consisting of complete data for $N_o<N$ units  to estimate the factors, and a \textsc{wide} block  in  which data for all units are available over $T_o<T$ periods to estimate the factor loadings. To align the space spanned by the estimated factors and loadings, one then estimates a  new rotation matrix by regressing  the $N_o\times r$ matrix of  loadings estimated from the  \textsc{tall} block on a sub-block of $N\times r$ matrix of loadings  estimated from the \textsc{wide} block. The procedure  produces    consistent and asymptotically normal estimates without iteration.  Re-estimation using the completed (imputed) data  will improve efficiency of the estimates in the balance block where the \textsc{tall} and \textsc{wide} blocks intersect to the fastest rate possible, which is the rate obtained in the complete data case.

\begin{figure}[ht]

\caption{Missing Data with a Less Structured Pattern}
\label{fig:fig2}
{\footnotesize
	\begin{center}
		
\begin{tabular}{cc}		
Reverse Monotone Missing  & \quad Heterngeneous Missing \\

\begin{tabular}{ c|c |c | c | c |c| c |  c|c|c|}

      \multicolumn{1}{r}{} &
      \multicolumn{1}{c}{$X_{1}$}  &
      \multicolumn{1}{c}{$\ldots$}
         &\multicolumn{1}{c}{$X_{N_o}$} &
      \multicolumn{1}{c}{$X_{N_{o+1}}$} &
       \multicolumn{1}{c}{$\ldots$} &
       \multicolumn{1}{c}{$X_{N-1}$} &
        \multicolumn{1}{c}{$X_{N}$} \\ \cline{2-8}

	&	&    &    &    & x   &  x  & x    \\  \cline{2-8}
	&	&    &    &   &    &   x  & x    \\  \cline{2-8}
	&	 &    &    &    &    &    & x          \\ \cline{2-8}
	&	&    &    &    &    &   & x         \\ \cline{2-8}
	&	 &    &    &     &    & &             \\ \cline{2-8}
	&	 &    &    &     &    & &            \\ \cline{2-8}
		\end{tabular}
		& \quad\quad

\begin{tabular}{ c|c |c | c | c |c| c | c|c|}
  \multicolumn{1}{r}{} &
      \multicolumn{1}{c}{$X_{1}$}
      &\multicolumn{1}{c}{$\ldots$}
     &\multicolumn{1}{c}{$X_{N_o}$} &
      \multicolumn{1}{c}{$X_{N_{o+1}}$} &
       \multicolumn{1}{c}{$\ldots$} &
       \multicolumn{1}{c}{$X_{N-1}$} &
        \multicolumn{1}{c}{$X_{N}$} \\ \cline{2-8}

&    &   &     &    & &   &   \\  \cline{2-8}
&    &   &     &    & &      &   \\  \cline{2-8}
&    &   &    &     & &      &  x  \\  \cline{2-8}
&    &   &    &     & &      & x       \\ \cline{2-8}
&    &    &    &    & x &   x   &   \\  \cline{2-8}		
&    &    &    &    & &      & x  \\  \cline{2-8}		
		\end{tabular}
	\end{tabular}
	\end{center}
}
\end{figure}

The \textsc{tw} algorithm takes  as given  that the \textsc{tall} and \textsc{wide}  blocks are 'big enough' for consistent estimation.
Because the size of the missing data block is defined by number of time
periods  and  units  with complete cases,    the size of the missing block is the same for both examples  in Figure \ref{fig:fig1}. The algorithm uses information available efficiently when the missing pattern is homogeneous.  But for the example in the right panel,   more  information could have been used.

More concerning are situations when  a few series significantly reduce $N_o$ and $T_o$.
As an example, consider  Figure \ref{fig:fig2}. The left panel shows a case of reverse monotone missing. In multi-country panels, this pattern will arise if data for less developed countries are only available at a later date than developed countries.   The right panel shows a case where the sample size  is much smaller for unit $N$ than for units $N-1$ and $N-2$. In both examples, the missing block   can contain  many observed values.


\section{A Projection Based  Procedure for Imputing $X$}


In this section, we will develop a new  procedure  to obtain consistent estimates without iteration as in \textsc{tw}, but can accommodate  flexible  missing data patterns while  making more efficient use information available.   The point of departure is to partition the $T\times N$  matrix $X$ in a different way.
We assume that there are $N_o$ series with no missing values so there is  a $T\times N_o$ block  labeled \textsc{tall}, and a $T\times (N-N_o)$ block of partially observed data labeled \textsc{ incomplete}. There is no need to reorganize the data in practice, but Figure \ref{fig:figtp} shows the data with   the $N_o$  variables  ordered first    to help visualize the idea.
\begin{figure}[ht]
\caption{The Tall-Incomplete Representation of the Data}
{\small
\begin{center}
\begin{tikzpicture}
  \matrix[
  matrix of math nodes,
  row sep=.5ex,
  column sep=4ex,
  left delimiter=(,right delimiter=),
  nodes={text width=.75em, text height=1.75ex, text depth=.5ex, align=center}
  ] (m)
  {
    \x   & \x  & \x & \x  & \x & \x &  \x & \x\\
    \x   & \x  & \x & \x  & \x & \x  & \x & \x\\
    \x   & \x   & \x & \x & \x & \x &   \x & \x\\
    \x   & \text{tall}   & \x  & \x &  \text{incomplete} &   & \y & \x\\
    \x   &  T\times N_o  & \x  & \x & T\times (N-N_o) &  &\y & \y\\
    \x   &  \x  & \x  & \x &\y & \x  &\y & \x\\
    \x   &  \x  & \x  & \x & \y & \y &\x & \y \\
    \x   &  \x  & \x  & \y & \y & \y &\y & \y \\
    };
    \begin{scope}[on background layer]
           \node[fit=(m-1-1)(m-8-8),
            fill=blue!40, rounded corners] {};
            \node[fit=(m-4-1)(m-8-3), fill=blue!15,rounded corners] {};
            \node[fit=(m-1-1)(m-3-3), fill=blue!15,rounded corners] {};
    \end{scope}
  \end{tikzpicture}
\end{center}
}
\label{fig:figtp}
\end{figure}

 It will be helpful to  define two locator sets  as follows:

\begin{center}
\begin{tabular}{ll}
$J^t$ &:= $\{j: X_{jt} \text{ observed,  i.e. units  with data in period t} \}$\\
$J_i$ &:= $\{s:  X_{is} \text{ observed, i.e.  periods with data for unit i} \}$
\end{tabular}
\end{center}
Let $N_{o_t}$ be the number of units observed at time  $t$ and $T_{o_i}$  be the number of rows observed for series $i$.
Then $J^t$ is a $N_{o_t}\times 1$ vector  that keeps track of the units observed at $t$, and $J_i$ is a $T_{o_i}\times 1$ vector that keeps track of the periods that unit $i$ is observed. For a set $\mathcal A$, let $|\mathcal A|$ be its cardinality.  In this notation,  $N_o=| \cap_t J^t|$,   $T_o=| \cap_i J_i|$, and  $T_{o_i}=T$ for every  unit in \textsc{tall}.

\paragraph{Algorithm Tall-Project (TP):}

\begin{itemize}
  \item[i.] Estimate  the $T\times r$ matrix  $\tilde F$  from the \textsc{tall}\; block by APC and let $\tilde F_{t}$ of dimension $r\times 1$ be the $t$-th row of $\tilde F$.
\item[ii.] For each $i$, regress the $T_{o_i}\times 1$ vector consisting of the observed values of $X_i$ on  the corresponding $T_{o_i}\times r$ submatrix of $\tilde F$  to obtain  $\tilde \Lambda_i$.
\item[iii.] For each $(i,t)$, let $\tilde C_{it}=\tilde F_{t}^\prime \tilde\Lambda_i$ and $\tilde e_{it}=\tilde X_{it}-\tilde C_{it}$ where
\[
  \tilde X_{it}=
  \begin{cases} X_{it}
    & \text{if }  \quad X_{it} \text{ observed }\\
    \tilde C_{it} &  \text{if } \quad  X_{it} \quad \text{missing}.
  \end{cases}
   \]

\end{itemize}
Step (i) of  \textsc{tp} is the same as  \textsc{tw} since both procedures estimate $F$ from the \textsc{tall} block.  If such a block does not exist, then neither \textsc{tw} or \textsc{tp} will be appropriate. Fortunately,  for macroeconomic panels,  a \tall\ block is often available. Algorithm \textsc{tp}  accommodates  staggered and irregular patterns of missingness  by  estimating $\Lambda_i$   using a customized sample for each series.  The two algorithms should be numerically identical if $ T_{o_i}=T_o$ for all $i$, but
since  $N_{o_t}\ge N_o$ and $T_{o_i}\ge T_o$,   \textsc{tp} will utilize more  information in general. The difference  can be significant if  $T_o$  is dictated by a few series with many missing observations. Note, however,  that
\textsc{tp} estimates the $H\Lambda_i$ matrix  directly, while \textsc{tw}    estimates $H$ and $\Lambda_i$ separately.  Furthermore, the number of factors in Algorithm \textsc{tp}    is determined by the \tall\; block so  Algorithm \textsc{tw} is more flexible in this regard.

We will need  additional assumptions to analyze the properties of \textsc{tp}.

\paragraph{Assumption B:} $\frac{\sqrt{N}}{\min\{N_o,T_o\}}\rightarrow 0$ and $  \frac{\sqrt{T}}{\min\{N_o,T_o\}}\rightarrow 0$ as $N\rightarrow\infty$ and $T\rightarrow\infty$.

\paragraph{Assumption C:}    There exists  $M<\infty$ such that for all $N$ and $t$,
 $ \frac {N_{o_t}(N - N_{o_t})} {N N_o} \le M$.

\paragraph{Assumption D:}

For each $i$, $\Big(\frac 1 {T-T_{\oi}}  \sum\limits_{s\notin J_i} F_s^0 F_s^{0\prime}\Big) \Big( \frac1 {T_{\oi}} \sum\limits_{s\in J_i} F_s^0 F_s^{0\prime} \Big)^{-1}
\pconv I_r,$
 $\frac 1 {\sqrt{T_{o_i}}  } \sum\limits_{s\in J_i}  F_s^0 e_{is} \dconv N(0,\Phi_i).$
For each $t$,  $\Big(\frac 1 {N-N_{o_t}} \sum_{k\notin J^t} \Lambda_k^0\Lambda_k^{0\prime}\Big) \bigg(\frac{1}{N_o} \sum_{k \in \cap_s J^s} \Lambda_k^{0\prime}\Lambda_k^0\bigg)^{-1}\pconv I_r$, $\frac{1}{\sqrt{N_{o_t}}} \sum_{k \in J^t}  \Lambda_k^0 e_{kt}\dconv N(0,\Gamma_t)$.

\bigskip

Assumption B essentially puts a lower bound on the observed number of rows and columns. Assumption C puts an upper bound on the extent of missing values at any $t$ and can be understood as a noise-to-signal constraint. Assumption C allows $N_{o_t}/N\rightarrow 0$, and especially $N_o/N\rightarrow 0$. For example,
if $N_{o_t}=N_o$, then the ratio is bounded by 1.
For the   central limit theorems in Assumption D to hold, $T_{o_i}$ should be independent of $(F^0_{s}, e_s)$, and $N_{o_t}$  independent of $(\Lambda_k^0,e_k)$.  The assumption is satisfied if the  missing data are unrelated to the intrinsic properties of the series.  For example, missing data due to mergers are allowed if the events are unrelated to $F_t^0$. However,    missing data due to bankruptcies would not be allowed if the bankruptcy probability  depends on $F_t^0$.
These  conditions are  stronger  than   stationarity and need to be justified on  an application by application basis. All factor-based imputation procedure requires a similar assumption because without some commonality between the observed and missing units, imputation would not be possible.

\begin{lemma}
  \label{lem:tp} (First Pass Estimation) Let $N_o$ be the number of units in the \textsc{tall} block and $T_{o_i}$ be number of periods that unit $i$ is observed. Let $H_{tall}$ be a rotation matrix based on the \textsc{tall} block of the data.  Under Assumptions A-D,  the \textsc{tp} estimates $(\tilde F,\tilde \Lambda)$ have  the following properties:
 \begin{itemize}
  \item[i.] $ \sqrt{N_{o}}(\tilde F_{t}- H_{tall}^\prime F^0_t)\dconv  N\bigg(0,\mathbb D_r^{-2}\mathbb Q_r \Gamma_t \mathbb Q_r^\prime \mathbb D_r^{-2}\bigg)$, for  $t \in [1,T]$, if $\frac{\sqrt{N_{o}}}{T}\rightarrow 0$

  \item[ii.] (a) $\sqrt{T} (\tilde\Lambda_{i}- H_{tall}^{-1} \Lambda_i^0)\dconv N\bigg(0, (\mathbb Q_r^\prime)^{-1} \Phi_i )\mathbb Q_r^{-1} \bigg) $, for $i\le N_o$, if $\frac{\sqrt{T}}{N_o}\rightarrow 0$ \\
  (b) $\sqrt{T_{o_i}} (\tilde\Lambda_{i}- H_{tall}^{-1} \Lambda_i^0)\dconv N\bigg(0,(\mathbb Q_r^\prime)^{-1} \Phi_i )\mathbb Q_r^{-1} \bigg) $, for $i>N_o$, if $\frac{\sqrt{T_{o_i}} } {N_{o}}\rightarrow 0$
\item[iii.] Let $\tilde {\mathbb V}_{it}({ N_o, T_o})$ be a consistent estimate of $ \mathbb V_{it}=     \frac{\delta^2_{ N_o, T_\oi}}{ N_o} \Lambda_i^{0\prime} \Sigma_{  \Lambda}^{-1}  \Gamma_t \Sigma_{\Lambda}^{-1}\Lambda^0_i+
     \frac{\delta^2_{ N_o, T_\oi}}{ T_o}F_t^{0\prime} \Sigma_{F}^{-1}  \Phi_i \Sigma_{F}^{-1}F^0_t$ where $\delta^2_{N_o,T_{o_i}}=\min(N_o,T_{o_i})$.
   Then
\[  \min(\sqrt{N_o},\sqrt{T_{o_i}})\bigg(\frac{\tilde C_{it}-C_{it}^0}
  {\sqrt{\tilde{\mathbb V}_{it}(N_o,T_{o_i})}}\bigg)
   \dconv  N(0,1).\]

\end{itemize}

\end{lemma}
Part $i$ and  $ii(a) $  are implied by Lemma \ref{lem:bai-03} because $\tilde F_t$ and $\tilde \Lambda_i$ are estimated from the \textsc{tall}  block, which is a complete matrix of $T \times N_o$ dimension. Result (b) of part ii arises  because  the factor loadings for unit $i$ in the \textsc{incomplete} block are estimated from a regression with $T_{o_i}$ observations. Part (iii) shows  that each $\tilde C_{it}$ has its own convergence rate that depends on $T_{o_i}$, the number of observed entries  in $X_i$. The result
 is  based on  the
fact for  those with  $X_{it}$ that are missing,
\begin{eqnarray}
 \tilde C_{it}-C_{it}^0&=& \Lambda_i^{0\prime} \bigg(\frac{1}{N_o} \sum_{k \in \cap_s J^s} \Lambda_k^0\Lambda_k^{0\prime}\bigg)^{-1}\frac 1 {N_o} \sum_{k\in \cap_s J^s} \Lambda_k^0 e_{kt}
+F_t^{0\prime} \bigg(\frac{1}{T_{o_i}}
 \sum_{s \in J_i}  F_{s}^0 F_s^{0\prime}\bigg)^{-1}
\frac 1 {T_{o_i} } \sum_{s\in J_i} F_s^0 e_{is} +r_{it} \nonumber\\
&\stackrel{\Delta}{=}& u_{it}+v_{it}+r_{it}.
  \label{main-cit-cit}
\end{eqnarray}
 The error due to estimating $F$ from the \textsc{tall} block appears as the first term on the right hand side and is denoted $u_{it}$, while the error due to estimating $\Lambda_i$ by projections is summarized by the second term and is   denoted $v_{it}$. The term  $r_{it}=O_p(\delta^{-2}_{T_o,N_o})$ uniformly in $i$ and $t$ represents high order errors when estimating the factor and factor loadings. These quantities  depend on $N$ and $T$ but the notation is suppressed for simplicity. The representation (shown in the Appendix) is useful in understanding how $N$ and $T$ affect factor-based imputation.

\subsection{Re-estimation using $\tilde X$}
As $\tilde F$ is estimated using the \textsc{tall} block alone,
 re-estimation of the factors and the loadings from the completed (imputed) matrix $\tilde X$  appears desirable. But imputation error in $\tilde X$ must be taken into account.  Using (\ref{main-cit-cit}),  we have
\begin{subequations}
\begin{align}
\tilde   X_{it}&=  \Lambda_i^{0\prime}F_t^0 +e_{it} , \quad  & \text{if} \quad X_{it}  \quad \text{observed} \label{eq:Xtilde-obs}\\
 \tilde X_{it}&= \Lambda_i^{0\prime}F_t^0 + u_{it}+v_{it} +r_{it}\quad & \text{if} \quad X_{it} \quad \text{missing}. \label{eq:Xtilde-miss}
\end{align}
\end{subequations}
 A quantity that will play a role in APC estimation from $\tilde X$ is
 \begin{eqnarray*}
 B^i_t =\begin{cases} \frac {N_{o_t}} N I_r+ \frac {N_{o_t}} N \Big(\frac{N-N_{o_t}}{\No} \Big)\Big(\frac 1 {N-N_{o_t}} \sum_{k\notin J^t}^N \Lambda_k^0\Lambda_k^{0\prime}\Big) \bigg(\frac{1}{N_o} \sum_{k \in \cap_s J^s} \Lambda_k^{0}\Lambda_k^{0\prime}\bigg)^{-1} \quad & i\in J^1\cap\cdots \cap J^T  \\
\frac {N_{o_t}} N I_r \quad & i \in J^t\setminus (J^1\cap\cdots \cap J^T)
\end{cases}
\end{eqnarray*}
 Under Assumption C,   $B^i_{t}$ is bounded. For units in the tall block (i.e., $i \in \cap_s J^s)$,  $B_t^i$ is roughly inflated from $\frac {N_{o_t}} N I_r$ by   $N_{o_t}(N-N_{o_t})/(NN_o)$ which can be thought of as the noise to signal ratio..

\begin{proposition} \label{prop:hatX-coro1}
  Let $(\tilde F^+,\tilde\Lambda^+)=(\sqrt{T} \tilde U_r,\sqrt{N} \tilde V_r\tilde D_r)$ where $\tilde U_r,\tilde V_r$ are the $r$ left and right singular vectors of $\tilde X$. Let $H^+$ be a rotation matrix.
  Under Assumptions $A-D$,
     the factors and loadings  estimated from $\tilde X$ have the following properties:
\begin{itemize}
  \item[i]  $\sqrt{ N_{o_t}  }(\tilde F^+_t-H^{+^\prime} F^0_t)\dconv N(0, \mathbb D_r^{-2} \mathbb Q_r \Gamma_t^* \mathbb Q_r^\prime \mathbb D_r^{-2})$
  \item[ii]  $\sqrt{ T_{o_i}}(\tilde \Lambda_i^+-(H^+)^{-1} \Lambda_i^0)\dconv N(0, (\mathbb Q_r^\prime)^{-1} \Phi_i (\mathbb Q_r)^{-1}))$

  \item[iii] Suppose that $e_{it}$ is  cross-sectionally uncorrelated. Let $\tilde C_{it}^+=\tilde \Lambda_i^{+\prime}\tilde F_t^+$.
Then, for all $(i,t)$.
\[ \min(\sqrt{N_{o_t}},\sqrt{T_{o_i}})\bigg(\frac{\tilde C_{it}^+-C_{it}^0}
  {\sqrt{\tilde{\mathbb V}^+_{it}(N_{o_t},T_{o_i}) }} \bigg)
   \dconv N(0,1) \]
where  $\tilde {\mathbb V}^+_{it}({ N_{o_t}, T_{o_i}})$ is a consistent estimate of
\[ \mathbb V_{it}^+=      \frac{\delta^2_{ N_{o_t} T_{o_i}}}{ N_{o_t}} \Lambda_i^{0\prime} \Sigma_{  \Lambda}^{-1}  \Gamma_t^* \Sigma_{\Lambda}^{-1}\Lambda^0_i+
     \frac{\delta^2_{ N_{o_t} T_{o_i}}}{ T_{o_i}}F_t^{0\prime} \Sigma_{F}^{-1}  \Phi_i \Sigma_{F}^{-1}F^0_t,\] $\delta_{N_{o_t},T_{o_i}}^2=\min(N_{o_t},T_{o_i})$ and
   $\Gamma_t^*= \text{plim} \frac 1 {N_{o_t}} \sum_{i \in J^t} B^i_{t} \Lambda_i^0 \Lambda_i^{0\pr} B^i_{t}\pr e_{it}^2$.
\end{itemize}
\end{proposition}

 The  proposition establishes the convergence rate of the  factors and loadings constructed from  $\tilde X$.
There are three changes due to re-estimation. First,   there is  only one (instead of many) rotation matrices for the factor loadings.  This is a consequence of the fact that the factors are now estimated from $\tilde X$ in its entirety, instead of sub-blocks.  Second, re-estimation generates efficiency gains.
 The convergence rate for $\tilde F_t^+$ is improved from $\sqrt{N_o}$ to $\sqrt{N_{o_t}}$ or to $\sqrt{N}$. Though the  rate for $\tilde \Lambda^+$ is unchanged,
 the convergence rate of $\tilde C^+_{it}$  is now $\min(\sqrt{N_{o_t}},\sqrt{T_{o_i}})$ instead of $\min(\sqrt{N_o},\sqrt{T_{o_i}})$. If $N_{o_t}=N$ and $T_{o_i}=T$ for a given pair $(i,t)$, the convergence rate for $\tilde C_{it}^+$ is  $\min(\sqrt{N},\sqrt{T})$, the best rate possible (the same as in complete data). Third, the asymptotic variance of $\tilde C_{it}^+$ depends on the $r\times r$ matrix $B_{t}^i$. From this result,  a 95\% confidence interval for $C^0_{it}$ of $\tilde C^+_{it} \pm 1.96 \tilde {\text{se}}(\tilde C^+_{it})$ where $\tilde{ \text{se}}(\tilde C^+_{it})=\sqrt{\frac{\tilde {\mathbb V^+_{it}}(N_{o_t},T_{o_i})}{\min(N_{o_t},T_{o_i})}}$.  The prediction interval for a missing $X_{it}$  is $\tilde X^+_{it}\pm 1.96 \tilde {\text{se}}(\tilde X^+_{it})$  where $\tilde{ \text{se}}(\tilde X^+_{it})=(\tilde \sigma_{ei}^2+\tilde{ \text{se}}(\tilde C^+_{it})^2)^{1/2}$, $\tilde\sigma_{ei}^2=\frac{1}{T_{o_i}}\sum_{s \in J_i} \tilde e_{is}^2$  is an estimate of the variance of $e_{it}$.

\subsection{Implications for Factor Augmented Regressions}

Consider  the infeasible regression with observed  covariates  $W_t$ and a latent predictor $F_t$:
\begin{eqnarray}
y_{t+h}&=& \alpha^\prime F_t+\beta^\prime W_t+\epsilon_{t+h} \label{eq:favar}\\
&=&\delta^\prime z_t+\epsilon_{t+h} \nonumber
\end{eqnarray}
where $z_t=(F_t^\prime, W_t^\prime)^\prime)$ and $\delta=(\alpha^\prime,\beta^\prime)^\prime$. Suppose that $F_t$ is replaced by an estimate based on  a small number of predictors.
It is known that  in general, the sampling uncertainty in a generated regressor  will inflate  standard errors in subsequent regressions.  A useful result in \citet{baing-ecta:06} is that  if  $F$ is estimated by APC  from a completely observed panel  $X$, then $\tilde F$  can be used in a second step regression without the need for standard error adjustments    if certain conditions on the sample size are satisfied.  Our foregoing analysis suggests that even if $X$ has missing values,  $\tilde F^+$ estimated by \textsc{tp}  is still consistent for the space spanned by $F$ albeit at a slower convergence rate.

\begin{lemma}
Suppose  Proposition \ref{prop:hatX-coro1} holds. Then (i)
$ \frac{1}{T} \sum_{t=1}^T \|\|\tilde F_t^+-\mathbf H_{NT}^{\prime}F_t\|\|^2=O_p(\min[N_o,T_o]^{-1}).$
 Let $\hat z_t=(\hat F_t^{+\prime}, W_t^\prime)^\prime$  be used in place of $z_t$ in the factor-augmented regression (\ref{eq:favar}), and
 denote $\delta^0=(\alpha^\prime \mathbf H_{NT}^{\prime -1}   ,\beta^\prime)^\prime$. If $\sqrt{T}/N_o\rightarrow 0$ and $\sqrt{T}/{T_o}\rightarrow 0$,  then (ii)
\begin{equation}
 \sqrt{T}(\hat\delta-\delta^0)\dconv N(0,\mathbf J^{'^{-1}} \mathbf \Sigma_{zz}^{-1}\mathbf \Sigma_{zz,\epsilon}\mathbf \Sigma_{zz}^{-1}\mathbf J^{-1}).
\label{eq:favar2}
\end{equation}
where $\mathbf J$ is the probability limit of $\mathbf J_{NT}=\text{diag}(\mathbf H_{NT}', \mathbf I_{\text{dim}(W)})$.
\end{lemma}

Part (i) is based on Lemma 3 of \citet{baing-miss:jasa} obtained for \textsc{tw} which can be used as a worse case rate for \textsc{tp} since $T_o$ in \textsc{tw} cannot exceed $\min_i T_{oi}$ in \textsc{tp}. A consequence of (i) is that
\begin{eqnarray*}
 \frac{1}{\sqrt{T}}\sum_{t=1}^T  \hat z_t(F_t^{0\prime}\mathbf H_{NT}- \tilde F_t^{+\prime})
&=&O_p\bigg(\frac{\sqrt{T}}{\min(N_o,T_o)}\bigg),\\
\frac{1}{\sqrt{T}}\sum_{t=1}^T \epsilon_{t+h} \tilde F_t^+&=&\frac{1}{\sqrt{T}} \sum_{t=1}^T \epsilon_{t+h} \mathbf H_{NT}'F^0_t +\frac{1}{\sqrt{T}}\sum_{t=1}^T (\tilde F^+_t-\mathbf H_{NT}' F_t^0) \epsilon_{t+h}\\
&=&\frac{1}{\sqrt{T}}\sum_{t=1}^T \epsilon_{t+h} \mathbf H_{NT}' F_t^0+O_p\bigg(\frac{\sqrt{T}}{\min(N_o,T_o)}\bigg).
\end{eqnarray*}
These results are used to  obtain  (ii). In particular,  let $\mathbf S_{\hat z\hat z}=\frac{1}{T}\sum_{t=1}^T \hat z_t\hat z_t'$. Then
\begin{eqnarray*}\sqrt{T}( \hat\delta-\delta_0)&=& \mathbf S_{\hat z'\hat z}^{-1}\frac{1}{\sqrt{T}} \sum_{t=1}^T  \hat z_t\epsilon_{t+h}
+\mathbf S_{\hat z'\hat z}^{-1}\frac{1}{\sqrt{T}}\sum_{t=1}^T \hat z_t\alpha' \mathbf H_{NT}^{'^{-1}} (\mathbf H_{NT}'F_t^0 -\tilde F_t^+)'\\
&=& \mathbf S_{\hat z'\hat z}^{-1}  \mathbf J_{NT}\frac{1}{\sqrt{T}}\sum_{t=1}^T z_t\epsilon_{t+h} +o_p(1)
\end{eqnarray*}
  provided $\frac{\sqrt{T}}{N_o}\rightarrow 0$,  $\frac{\sqrt{T}}{T_o}\rightarrow 0$. If, in addition, $\frac{1}{\sqrt{T}} \sum_{t=1}^T z_t\epsilon_{t+h}\dconv N(0,\mathbf \Sigma_{zz,\epsilon})$, the asymptotic variance of $\hat\delta$  is of the sandwich form $\mathbf  A^{-1} \mathbf B \mathbf A^{-1}$ with   $\mathbf A=\mathbf J\mathbf \Sigma_{zz}\mathbf J$ and $\mathbf B = \mathbf J\mathbf \Sigma_{zz,\epsilon}\mathbf J'$, where   $\mathbf J$ is the probability limit of $\mathbf J_{NT}$. Simplifying yields (ii).
The result implies that  as  in \citet{baing-ecta:06}, the estimated factors can be treated as though $F$ were observed albeit under more stringent conditions than the complete data case which only requires that $\sqrt{T}/N\rightarrow 0$. For macroeconomic analysis when the factors are estimated from panels  with large $T_o$ and $N_o$,  one can still expect    $\hat\delta$ to be precisely  estimated up to a rotation.

\section{Factor Based Estimation of Covariance Matrices}
 Consider portfolio analysis where $X$ is a matrix of returns and $\Sigma_X$ is its covariance.  The population weights of   a minimum variance portfolio are    given by  $  w_p  = \frac{ \boldsymbol{ \Sigma_X^{-1} 1}}{ \boldsymbol{ 1\prime \Sigma_X^{-1} 1} }$
where $\boldsymbol{1}$ is the unit vector.  If $X$ was completely observed and  $x_i=X_i-\bar X_i$, the  sample moments (SM) estimator of the $N\times N$ matrix is
\begin{align*}
\hat {\mathbf \Sigma}_{X} =& \frac{1}{T} \sum_{t=1}^T  x_t x_t^\prime.\tag{SM}
\end{align*}
 It is well known  that the sample covariance matrix is singular when $N>T$ and has large sampling uncertainty when
 $N$ and $T$ are large.  Nonetheless, when the data admit a factor structure so that the decomposition   $\mathbf \Sigma_X=\mathbf \Lambda\mathbf \Sigma_F \Lambda^\prime+\mathbf \Sigma_e$ holds, a factor-based  covariance estimator can be defined as
\begin{align*}
\hat{\mathbf \Sigma}_{ X}=&\tilde{\mathbf \Lambda}\widehat{\mathbf \Sigma}_{\tilde F}\tilde{\mathbf \Lambda}'+\hat{\mathbf \Sigma}_{\tilde e}
\end{align*}
where  $\hat{\mathbf \Sigma}_{\tilde e}$ is the sample covariance  of $\tilde {\mathbf e} $, which is never exactly diagonal even when $\mathbf \Sigma_e$ is  diagonal.   As argued in \citet{chamberlain-rothschild}, an approximate factor model that allows for some correlation in the idiosyncratic errors  is usually a better characterization of returns data.   To distinguish a strict  from an approximate factor structure,  let $\mathbf \Psi_e$ be  a  diagonal matrix whose $i$-th entry is   $\mathbf E[e_{it}^2]$ and which can be consistently estimated by

\begin{equation} \label{eq:Psi-e}
\tilde {\mathbf \Psi}_{\tilde e} =\text{diag}\bigg(\frac{1}{T}\sum_{t=1}^T   \tilde e_{it}^2,i=1,2,...,N\bigg).
\end{equation}
The strict-factor (SF) covariance estimator is defined as
\[
\mathbf {\tilde  \Sigma}_{ X}=
   \tilde \Lambda \tilde{\mathbf \Sigma}_F \tilde \Lambda^\prime+ \tilde { \mathbf \Psi}_{\tilde e}
\tag{SF}
\]
Consistency of this estimator is studied by \citet{fan:11}, among others, for large $N$ and large $T$, assuming that $X$ is completely observed.

We now turn to  the incomplete data case. Missing data presents a challenge for risk management  because the  portfolio weights depend on $\mathbf \Sigma_X$.  Variable deletion is not an option when the variances and covariances of the missing variables are the objects of interest. Practitioners often resort to  listwise deletion of the sample data, deleting the entire record  from the analysis if any single value is missing.   The  covariance estimates can be unstable when the reduction in the sample  size is sufficiently large.  Though it  is possible  to calculate the sample covariance matrix with pairwise-complete observations, there is no guarantee that the resulting matrix will be  positive definite.   This is problematic because a singular covariance matrix must have at least one eigenvalue that is zero, implying that it would be possible to construct an  eigenportfolio  that has zero volatility (risk) by using the corresponding  eigenvector as the weighs, making it possible to have an infinite ex-ante Sharpe ratio. Not only is this unrealistic,  the hedges implied by the eigenportfolio weights are spurious and would fail out of sample.  For these reasons,   singular covariance matrices are of little use in   portfolio construction in practice. Most  risk applications require an estimate of a  non-singular covariance matrix. The EM algorithm is one possibility, but it is likelihood based and requires parametric assumptions.

Unfortunately, successful factor-based imputation of the level of the data is not enough for precise estimation of their  covariances. Even if the strict factor structure is correctly imposed,
 both the sample-covariance estimator and factor-based  covariance estimator based on the imputed data will not be  consistent.   The problem with both estimators is that  $\tilde e_{it}$ is set to zero when $X_{it}$ is not observed, so the sample variance of series $i$ is biased whenever the series has missing values. We can
  replace $\tilde{\mathbf \Psi}_{\tilde e}$ in (\ref{eq:Psi-e}) by
\[
\hat {\mathbf \Psi}_{\tilde e} =\text{diag}\bigg(\frac{1}{T_{o_i}}\sum_{t\in J_i}   \tilde e_{it}^2,i=1,2,...,N\bigg)
\]
to  recognize that some $\tilde e_{it}$ are zero, and the estimate $\hat\Psi_{\hat e,ii}=\frac{1}{T_{o_i}}\sum_{t\in J_i}   \tilde e_{it}^2$ is  consistent for  $\mathbf \Psi_{e,ii}$ if $T_{\oi}$ is sufficiently large.  Define the strict-factor adjusted (SFA) covariance matrix estimator  as
\[
\mathbf {\hat  \Sigma}_{\tilde X}=
   \tilde \Lambda \tilde{\mathbf \Sigma}_F \tilde \Lambda^\prime+ \hat { \mathbf \Psi}_{\tilde e}
\tag{SFA}
\]

\begin{lemma} \label{lemma-S-S}
Let $\hat {\mathbf \Sigma}_{\tilde X}$ be the \textsc{sfa} estimate of $\mathbf \Sigma_X$.
  Under Assumptions A, B,  exponential tail distribution for $e_{it}$, and that  $\Sigma_e$ is a diagonal matrix,
   \[ \| \hat {\mathbf \Sigma}_{\tilde  X}-\mathbf \Sigma_X\|_\infty =\max_{1\le {i,j}\le N} |\hat \Sigma_{\tilde X,ij}-\Sigma_{X,ij}| =o_p(1). \]
where  for a matrix $\mathbf A$ with  $A_{ij}$ as its $(i,j)$ entry,  $\|\mathbf A\|_\infty=\max_i \max_j |A_{ij}| $.
\end{lemma}
The proof is given in the Appendix.

\subsection{SM  Estimation and Double Imputation}

This subsection considers a sample covariance estimator as an alternative to the SFA covariance estimator just described. Motivated by the fact that  the variance of an imputed variable is downward biased,  we
 use a second imputation  to rectify this problem.
\paragraph{Algorithm Residual Overlay}

 Let $\tilde {e} = \tilde X - \tilde C$ where $\tilde X$ is produced by Algorithm \textsc{tp}.
For  residual resampling scheme $j$ ($j=1,2,3,4)$ to be discussed below, repeat for $s =1,\ldots S$:

\begin{itemize}
\item[a] For each $i$ with missing data, replace those
 $\tilde e_{it}=0$ with  a   $\tilde e_{it}(s,j)$ randomly sampled  from those non-zero $\tilde e_{it}$ associated with observed $X_{it}$.
\item[b.] Define
$\hat{e}_{it}(s,j) =
\begin{cases}
\tilde e_{it} & \text{ if }  X_{it} \quad \text{observed} \\
 \tilde{e}_{it}(s,j) &\text{ if } X_{it}  \text { is missing}
\end{cases}$.
\item[c.] Let $\hat X_{it}(s,j)=\tilde\Lambda_i^\prime\tilde F_t+\hat e_{it}(s,j)$ and
 $\hat x_{it}(s,j)=\hat X_{it}(s,j)-\bar{\hat  X_i}(s,j)$. Estimate the  covariance of $\hat X(s,j)$ as
$\hat{\mathbf \Sigma}_{\hat X}(s,j)=\frac{1}{T}\sum_{t=1}^T \hat x_t(s,j)\hat x_t(s,j)^\prime$.

\item[d.]  The $N\times N$  covariance estmator obtained from  $\tilde X$ imputed by TP and overlaid with residuals sampled using  scheme $j$ is
\[\bar {\bf \Sigma}_{\tilde  X}(j) = \frac{1}{S} \sum_{s=1}^{S} \hat{\bf \Sigma}_{\hat X}(s,j). \tag{SM+j}
\]

\end{itemize}
The term residual overlay is motivated by the fact that errors are added to $\tilde X$.
The algorithm  injects randomness  to $\tilde X_{it}$ whenever $X_{it}$ is missing. It uses stochastic simulations to compensate for the lost variability, an idea briefly considered  in \citet[Chapter 5]{enders:10} in a fixed $N$ or $T$ setting.  Even though each $\hat \Sigma_{\hat X}(s,j)$   is large in dimension and has rank $\min(N,T)$, in our experience,  the average $\bar{\bf \Sigma}_{\tilde X}$ is  always full rank, making it a viable alternative estimator in the case when $N>T$.

 The method of multiple imputation typically estimates the object of interset each time an imputed set of data is obtained, and produces  as output   the average over the multiple estimates. In contrast, we average the multiple imputed covariances to obtain  $\bar{\mathbf \Sigma}_{\hat X}$ and  compute the object of interest (for example, portfolio weights) from it.   Averaging has the advantage of  reducing sampling uncertainty especially when $N$ and $T$ are both large.

\paragraph{Step a: Four Sampling Schemes}
\begin{itemize}
 \item[-] (\textsc{sm1}) Let $u$ be  a $\sum_{i=1}^N T_{o_i}$  vector, obtained by stacking up all   $\tilde e_{it}$ associated with the observed $X_{it}$. Resample $\sum_{i=1}^N (T-T_{\oi})$ observations from $u$ with replacement and randomly assign them to $\hat e _{it}$ whenever  $\tilde e_{it}=0$.
\item[-] (\textsc{sm2}) For each $i$, let $u_i$ be  $T_{\oi}$ sub-vector of non-zero entries of  $\tilde e_i$. Sample  with replacement  $T-T_{\oi}$ errors from $u_i$  and assign them  to $\hat e_{it}$  whenever  $\tilde e_{it}=0$.
 \item[-] (\textsc{sm3})  Compute  $\hat\sigma_u$ from the $\sum_{i=1}^N T_{o_i}$ vector of
 non-zero estimated errors as in Method 1. If $\tilde e_{it}=0$, replace it by $\hat e_{it} :=  u_{it}\hat \sigma_u$, where $u_{it}\sim N(0,1)$.
 \item[-] (\textsc{sm4})  Compute $\hat\sigma_{u,i}$ from the $ T_{o_i}$
 non-zero estimated errors as in Method 2. If $\tilde e_{it}=0$, replace it by $\hat e_{it} :=  u_{it}\hat \sigma_{u.i}$, where $u_{it}\sim N(0,1)$.

\end{itemize}

All four methods  assume that the returns data are covariance stationary.\footnote{If the variances and correlations  change over time,  the covariance matrix can be calculated by estimating  GARCH models, for example.} Methods \textsc{sm1}  and \textsc{sm2} are non-parametric, while method \textsc{sm3} and \textsc{sm4} calibrate the first two moments  to the estimated errors associated with the observed data. Methods \textsc{sm1} and \textsc{sm3} are better suited for homogeneous data and since $\hat  e_{it}$ are sampled from  the stacked up vector of estimated errors. Methods \textsc{sm2} and \textsc{sm4}  are better suited for heterogeneous data as they sample from the  errors of the corresponding series. Methods \textsc{sm1} and \textsc{sm2} do not make distributional assumptions about the errors.
 As many financial time series tend to have fat tails and skewed distributions, other parametric distributions can be used in place of the normal distribution in \textsc{sm3} and \textsc{sm4}. The four sampling schemes do not account for cross-correlation in the errors,  which implicitly shrinks $\mathbf \Sigma_e$ towards zero.  This can be desirable when estimating high dimensional covariance matrices.

Methods \textsc{sm}+1, \textsc{sm}+2, \textsc{sm}+3, \textsc{sm}+4 are likewise defined with $\tilde e_{it}$ replaced by $\tilde e_{it}^+=\tilde X_{it}-\tilde F_t^{+\prime} \tilde \Lambda_i^+$,
where  $\tilde F^+$ and $\tilde \Lambda^+$ are obtained from the principal components of the imputed data $\tilde X$. In principle, we can also compute a SF estimator $
\hat{\mathbf \Sigma}_{ \hat X}(j)=\tilde{\mathbf \Lambda}\tilde{\mathbf \Sigma}_{ F}\tilde{\mathbf \Lambda}'+\frac{1}{S}\sum_{s=1}^S \hat{\mathbf \Psi}_{\hat e}(s,j) $
 with  $
\hat{\mathbf \Psi}_{\hat e_{ii}}(s)=\frac{1}{T} \sum_{t=1}^T\hat e_{it}^2(s,j)$, but there seems no advantage  over SFA which already gives a good estimate of the variance of $e_{it}$ when $T_{oi}$ is sufficiently large. This is indeed the case in simulations, and  hence not considered.



Though injecting noise to imputed values is not new, overlaying the noise to $\tilde X$ imputed using SM   appears to be new.  The resampling involves  bootstrap draws.  Bootstrapping large dimensional matrices is not a trivial exercise. Our set up is different, as we only need to bootstrap the noise corresponding to the missing data, which is much smaller in dimension. However, we have to pre-estimate the factors and the loadings, and  bootstrapping the factor estimates is also not a trivial problem as seen in \citet{goncalves-perron:20}. The proof of consistency of this estimator is thus quite delicate and is beyond the scope of this analysis. However, we can use simulations to evaluate if the idea holds promise to be worthy of further investigation.

\section{Simulations}


In this section, we consider  four experiments: the first and second assess the accuracy of the asymptotic approximations for $\tilde C_{it}$ and $\tilde C_{it}^+$ given in Lemma 2 and Proposition 1. The third  evaluates the residual overlay procedures   assuming a strict factor structure, while the fourth uses  observed SP 500 returns as complete data  to mimic an approximate factor structure.

\subsection{Finite Sample Properties for TP}

The first experiment  compares the performance of \textsc{tp} with and without re-estimation.  The design of the monte-carlo is identical to the one used in \citet{baing-miss:jasa}. Data are generated from $F\sim N(0,D_r)$ and $\Lambda\sim N(0,D_r)$ with $r=2$,  the diagonal entries in $D_r$ are equally spaced between 1 and $1/r$, and  $e_{it}\sim N(0,1)$.  For each replication, the error in estimating $C_{it}$ is computed for four locations of $(i,t)$.  As benchmark, we consider   the  infeasible case of complete  data  labeled \textsc{complete}. Also reported are results for the EM algorithm  in  \citet{stock-watson:handbook-16} that  uses the estimates from the balanced panel as initial values. The algorithm repeatedly regresses $X$ on $F$ and then $X$ on $\Lambda$ till convergence, but  the converged  estimates  may not be  mutually orthogonal.
 We consider three versions of each  estimator: one applied  to the raw data $X$,  one to the demeanend data,  one to the  standardized data. These are labled TP2, TP1, TP0 in the tables reported. The means and standard deviations are computed using the observations available for each series.

  Table \ref{tbl:tableTP} reports  the  root-mean-square-error for  four chosen $(i,t)$ pairs, one in each of the four blocks:  \textsc{tall}   is $T\times N_o$,  \textsc{wide}  is $T_o\times N$,  \textsc{bal}   is $T_o\times N_o$, and \textsc{miss}  is  $(T-T_o)\times (N-N_o)$,   With  $N=T=200$, results are reported  for four configurations of missingness. Evidently, the error   depends on  observability of $X_{it}$. The estimation error is  largest if $X_{it}$ is in the \textsc{miss} block and smallest when $X_{it}$ is in the \textsc{bal} block. For a given block, the estimator  errors are smaller when  the factors are re-estimated from $\tilde X$. This is consistent with the theory. Results for $(N,T)=(300,500)$ and $(500,300)$ are similar.

In the second experiment, we generate data from a model with two factors to assess the adequacy of the asymptotic approximations.   Two configurations of $(T,N) $ are considered:   $(300, 500)$  and $(500, 300)$ with $T_o=.4T$ and $N_o=.6N$. In both cases, about 15\% of the observations  are missing. The factors  and the loadings are  drawn from the standard normal distribution once and held fixed.  In each of the 5000 replications, a new batch of idiosyncratic errors are drawn from the normal distribution, so $X$ varies across replications but the locations of the missing values do not change. We then evaluate estimates of $C_{it}$ at four different  $(i,t)$ chosen in the neighborhood of $N_o,T_o$ so that they come from four blocks as defined above.

Row 1 of Table \ref{tbl:distSP1} reports the mean estimate of $C_{it}$ for  (i)  when  all data are observed, (ii) \textsc{tw0} (no re-estimation), (ii) \textsc{tw+} (re-estimation),  (iv) \textsc{tp0}, and (v) \textsc{tp+}. All five estimates are close to the true values, showing that \textsc{tw0} and \textsc{tp0} are consistent without iteration. The second row, which gives the standard deviation of the estimates in the monte-carlo, shows that the \textsc{tw+} and \textsc{tp+} estimates are slightly less variable than \textsc{tw} and \textsc{tp}, showing efficiency gains. The third row  gives the mean of the estimated asymptotic standard errors which are quite similar to row two, showing that the asymptotic approximation is quite accurate. The next two rows labeled $q_{05}$ and $q_{95}$ present the 5 and 95 percentage points  of the empirical distribution of  the standardized estimates. They are quite close to the quantiles from the normal distribution of $\pm 1.64$. The last row shows that coverage is generally satisfactory and reinforces the adequacy of the asymptotic normal approximation.
  In summary, the proposed  \textsc{tp} estimates are already consistent but the convergence rate of  $C_{it}$ depends on the position of  all $(i,t)$ as  given in Proposition \ref{prop:hatX-coro1}.  The updated  estimates make use of additional information and have improved statistical properties.

\subsection{Results for Covariance Estimation}
Having shown that the  properties of  $\tilde X$  are satisfactory, we proceed to evaluate the covariance estimates using economically meaningful benchmarks.
   From a $T \times N^*$ panel of complete data  $X^*$  treated as stock returns, we first calculate the  ``true"    $N^*\times N^*$ covariance matrix  $\bf \Sigma_{X^*}$. Using the  $N^*\times 1$ vector of  portfolio weights defined from $\bf \Sigma_{X^*}$, we  compute  $r^*_{pt}= \mathbf R^{*'}_t  w^*_{p} $ and  treat it as the ``true"  portfolio return at time $t$. We then randomly select $N$ stocks and  assume missing values  in the south-west block of the returns matrix  $\mathbf R^*$. Hence the data matrix for analysis is of dimension $(T,N)$.

For each method, the (in-sample) bias and root-mean-squared error are computed for  the following performance measures   over   $B=1000$ replications.
\begin{enumerate}
\item[i.]  \textsc{pvol}:  {\em Portfolio volatility}  defined  as
$ \textsc{riskp}     = \sqrt{\frac{1}{T}\sum_{t=1}^T (r_{pt}^* -\bar r_p^*)^2}$.

\item[ii]  \textsc{pvar}$_{\alpha}$: Portfolio value-at-risk  at confidence level $\alpha$   defined as
$Pr( r_p <  -VaR_{\alpha}) = 1- \alpha$.
\item[iii.] \textsc{call} options price  assuming that the current and strike price of each security are one-dollar, a risk-free rate of 2\%, and time to maturity of one year.

\item[iv] \textsc{var}: the variance  of returns
\item[v] \textsc{covar}: the covariance of returns.
\end{enumerate}

Of these measures, the first two are based on equally weighted portfolios while the last three are based on returns data. Portfolio volatility is often used as a risk measure when the benchmark  is cash.
 Portfolio value-at-risk is used as a measure of downside risk.\footnote{Portfolio VaR is generally quoted as a positive number (ie, as a loss).} For \textsc{call},   we price plain vanilla European call option prices written on SP500 stocks calculated using Black-Scholes formula. The call price, variance, and covariance measures  are  calculated only for returns with incomplete data. Since we have  $N_m$  series with missing values, we compute $N_m$ call prices and variances, and $ N_oN_m + (N_m)(N_m-1)/2$ covariances at each iteration.

The following notation is used Tables  \ref{tbl:simdata2a} to  \ref{tbl:sp500res2}.
\begin{itemize}\itemsep0em
\item \textsc{SM} (sample moments-based)  covariance estimators:
\begin{itemize}
\item[-] \textsc{sm0}: single imputation.
\item[-] \textsc{sm+0}: single imputation followed by  one re-estimation.
\item[-] \textsc{sm}j: double imputation after \textsc{sm0} using overlay method $j=1,\ldots,4$.
\item[-] \textsc{sm+}j: double imputation after \textsc{sm+}  using overlay method $j=1,\ldots,4$.
\end{itemize}
\item factor based covariance estimators: \textsc{sfa} and \textsc{sf+a}
\end{itemize}

\paragraph{Strict Factor Model:}   $X^*$ is simulated from a strict factor model:
\begin{align*}
& X^*_{it}=\lambda_i^{\prime} F_t + e_{it} \;\; , \;\;\;\; i=1,...,N^* \;\;, \;\; t=1,...,T  \\
& F_t \sim iid \; N(0,\sigma_F^2\mathbb{I}_r) \;\;,\;\; \lambda_i \sim iid \; N(0,\sigma_{\lambda}^2 \mathbb{I}_r)   \\
& e_{it} \sim iid \; N\left(0\;,\; \frac{1-R^2_i}{R_i^2}\sum_{q=1}^{r} \lambda_{iq}^2 \sigma_F^2 \right )
\end{align*}
\noindent where $\mathbb{I}_r$ is the $r \times r$ identity matrix, $\sigma_F^2$ and $\sigma_{\lambda}^2$ are the variances of each loading and factor, respectively, and $R_i^2$ is the percentage contribution of the systematic component to total variance (i.e. coefficient of determination) for series $i$. Note that setting the variance of $e_{it}$ in the way above guarantees that the coefficient of determination of each series is exactly equal to $R_i^2$. Since $r$ is assumed known,  we set $r=5$ without loss of generality. We  set $R^2_i$ to 0.6, $\sigma^2_{\lambda}$ to 1, and $\sigma^2_F$ to  $0.035$, respectively. These give  average  volatility of $9.6\%$, similar to monthly SP 500 returns of  $9.4\%$. The DGP abstracts from time varying loadings because the covariance estimators  and will be affected by omitted time variation  in the same way. Here, we focus on estimating the covariance  of $\tilde X$.

Table \ref{tbl:simdata2a} reports results over 1000 replications for  $(T,N)=(339, 100)$ with   15\% of the values missing (60\% of rows and 40\% of columns).    Whether we use \textsc{tw} or \textsc{tp} seems to make little difference for   \textsc{sm}j estimation. This can be due to the fact that either way,  the missing idiosyncratic errors are set to zero and does not make a big difference  for covariance estimation.  Re-estimation is, however,  beneficial for \textsc{sm} estimation especially when the first step is based on \textsc{tw}. This makes sense because \textsc{tw} uses a more restricted sample in the first step estimation and stands to gain more from re-estimation.  Of the four methods, \textsc{sm+2} and \textsc{sm+4} which resample from the non-zero $\tilde e_{it}$  of series $i$ alone have smaller bias and root-mean-squared error.

Turning to the factor-based estimators, since  the idiosyncratic errors are mutually uncorrelated by design,  we had expected \textsc{sfa} and \textsc{sfa+} which impose the strict factor structure to give better estimates than the \textsc{sm} counterparts. While \textsc{sfa+} is comparable to \textsc{sm2} and \textsc{sm4}, it is  inferior to \textsc{sm+2} and \textsc{sm+4}.   This  can be due to the fact that  our implementation of \textsc{sm} is an average of $S$ imputed covariances which has a noise reducing property. It is also possible that the \textsc{sm} estimators  impose sparsity through univariate re-sampling, and as mentioned above, residual cross-section dependence  can only come from the non-zero $\tilde e_{it}$  associated with the non-missing data. As these are estimates of  errors that are uncorrelated by design,  the \textsc{sm} estimators should return sample covariances that are approximately diagonal.

As robustness check, the top panel of Table \ref{tbl:simdata2b} presents results for 30\% missing (80\% of rows and 60\% of columns). The relative performance  of the estimation methods are similar across levels of missingness, and as expected, estmation accuracy of a given model is higher when the missingness is lower.
The results so far are presented for $N<T$. Results  for $T=200, N=250$ are shown in the bottom panel Table \ref{tbl:simdata2b}. Note that with this design, each $\tilde{\mathbf  \Sigma}_{\tilde X}(s,j)$ is verified to be singular. However,   the averaged estimator $\bar {\mathbf \Sigma}_{\tilde X}(j)$ is not. Similar to results in Table \ref{tbl:simdata2a}, \textsc{sm+2} and \textsc{sm+4} perform well and often better than \textsc{sf+a}.

\paragraph{Monte Carlo Calibrated to SP500 Returns}
The final exercise is a Monte-Carlo that  takes $X^*$  to be SP500  monthly returns between 1990 and 2018 to capture the approximate factor structure. The first factor in $X^*$  of dimension  $348\times 339$ explains over 26.2\% of the variations, the second explains 4.1\%. The next three factors explain 3.8, 2.7, and 2.1\% of the variation, respectively. As in the previous monte carlo, we random select $N=100$ stocks in each replication and set some values in the south-west block to  missing.

The results based on 1000 replications are reported in Table  \ref{tbl:sp500res2}.   Imputation leads to some reduction in bias and variance, but  the errors are significantly smaller than no adjustment upon comparing \textsc{sm}1-4   with \textsc{sm0}.  Though double imputation is generally better than single  imputation,  the precise gain depends on the risk measure and the estimation sample. As in the strict factor case, \textsc{tw} and \textsc{tp} give similar risk-performance errors.  Also as in the strict factor case,  \textsc{sfa+} is comparable to \textsc{sm+1} and \textsc{sm+3}, but has larger errors than \textsc{sm+2} and \text{sm+4}.  However,
 improvements from re-estimation both in terms of bias and RMSE are larger for this DGP that mimics  the approximate factor structure.

Overall, we find that  single imputation is strongly preferred over no imputation but is inferior to  double imputation. Ultimately, the appropriate method depends on the data generating process and we offer several alternatives worthy of consideration.
  Averaging the covariances of imputed data overlaid with resampled errors has two desirable effects that we had not anticipated: it  reduces noise, and the averaged estimator is full rank. This interesting finding and the role of resampling in covariance estimation is an interesting problem that warrants further investigation.

\section{Conclusion}
This paper provides three sets of results. The first is a \textsc{tp} algorithm that can consistently estimate the entire low rank matrix  without iteration and a distribution theory for the estimates is provided. The second result makes precise the conditions under which factors estimated from incomplete data can be treated as known in factor-augmented regressions. The third pertains to estimation of  covariances  from incomplete data. We consider different schemes to compensate for an omitted error in the level estimates.  Implications for using imputed factors in augmented regressions are also discussed.
\clearpage

\begin{table}[ht]
  \caption{Root Mean-Squared-Error of $\tilde C_{it}$ at four $(i,t)$ pairs}
\label{tbl:tableTP}
\begin{center}
\begin{tabular}{ll|lll|lll|lll|lll} \hline
 &     &(0) & (1) & (2) & (0) & (1) & (2) & (0) & (1) & (2) & (0) & (1) & (2)\\ \hline
case & $(i,t)$ in
 & \multicolumn{3}{c|}{COMPLETE}
 & \multicolumn{3}{c|}{TP} &
 \multicolumn{3}{c|}{TP+}  &

\multicolumn{3}{c}{EM}  \\ \hline
 1 &   tall ( 200, 200) & 0.26 & 0.25 & 0.23 & 0.29 & 0.29 & 0.29 & 0.27 & 0.27 & 0.25 & 0.28 & 0.27 & 0.25 \\
  1 &   wide ( 120, 200) & 0.26 & 0.26 & 0.24 & 0.32 & 0.31 & 0.31 & 0.29 & 0.29 & 0.25 & 0.29 & 0.29 & 0.26 \\
  1 &    bal ( 200, 120) & 0.26 & 0.25 & 0.22 & 0.29 & 0.29 & 0.29 & 0.26 & 0.25 & 0.23 & 0.26 & 0.25 & 0.22 \\
  1 &   miss ( 120, 120) & 0.26 & 0.26 & 0.23 & 0.32 & 0.31 & 0.31 & 0.31 & 0.30 & 0.27 & 0.31 & 0.31 & 0.27 \\
\hline
  2 &   tall (  80,  80) & 0.26 & 0.25 & 0.23 & 0.30 & 0.29 & 0.29 & 0.29 & 0.28 & 0.25 & 0.29 & 0.28 & 0.25 \\
  2 &   wide ( 200, 200) & 0.26 & 0.26 & 0.23 & 0.38 & 0.38 & 0.37 & 0.35 & 0.35 & 0.30 & 0.36 & 0.35 & 0.30 \\
  2 &    bal ( 120, 200) & 0.26 & 0.25 & 0.23 & 0.29 & 0.29 & 0.29 & 0.27 & 0.26 & 0.23 & 0.26 & 0.26 & 0.23 \\
  2 &   miss ( 200,  60) & 0.26 & 0.25 & 0.23 & 0.38 & 0.37 & 0.37 & 0.37 & 0.36 & 0.32 & 0.38 & 0.38 & 0.33 \\
\hline
  3 &   tall ( 120,  60) & 0.26 & 0.25 & 0.22 & 0.36 & 0.35 & 0.35 & 0.32 & 0.31 & 0.29 & 0.32 & 0.31 & 0.28 \\
  3 &   wide (  80, 140) & 0.26 & 0.25 & 0.23 & 0.39 & 0.38 & 0.37 & 0.29 & 0.29 & 0.26 & 0.30 & 0.29 & 0.26 \\
  3 &    bal ( 200, 200) & 0.26 & 0.26 & 0.23 & 0.38 & 0.36 & 0.36 & 0.27 & 0.26 & 0.24 & 0.27 & 0.26 & 0.23 \\
  3 &   miss (  60, 200) & 0.26 & 0.26 & 0.23 & 0.39 & 0.38 & 0.37 & 0.35 & 0.34 & 0.31 & 0.35 & 0.34 & 0.31 \\
\hline
  4 &   tall ( 200, 120) & 0.26 & 0.25 & 0.23 & 0.38 & 0.36 & 0.36 & 0.35 & 0.34 & 0.32 & 0.35 & 0.33 & 0.31 \\
  4 &   wide (  60, 120) & 0.26 & 0.25 & 0.23 & 0.46 & 0.45 & 0.44 & 0.38 & 0.37 & 0.32 & 0.38 & 0.37 & 0.32 \\
  4 &    bal ( 140,  80) & 0.26 & 0.26 & 0.23 & 0.37 & 0.36 & 0.36 & 0.30 & 0.28 & 0.25 & 0.27 & 0.27 & 0.24 \\
  4 &   miss ( 200, 200) & 0.26 & 0.26 & 0.23 & 0.45 & 0.44 & 0.42 & 0.42 & 0.42 & 0.37 & 0.44 & 0.43 & 0.37 \\

 \hline
  \end{tabular}
\end{center}
   \small{Note:
 DGP: The $T\times N$ data matrix $X$ is generated by  $X=F\Lambda^\prime+e$, $F\sim N(0,D_r)$, $\Lambda\sim (0,D_r)$ with $r=2$, $e\sim N(0,2.5)$, and $\diag(D_r)=[1;.5]$. $(\mathbb N,\mathbb T$) is the number of columns and rows in the block. Four configurations of missing data are considered.  Case 1 havs the smallest \textsc{miss} block and case 4 has the largest. Reported are the root-mean-squared error over 5000 replications.}

\end{table}


\begin{table}
\begin{center}
\caption{Finite Sample Properties of Selected $\tilde C_{it}$}
\label{tbl:distSP1}
\begin{tabular}{r|rrrrr|rrrrr}
&  \textsc{apc} & \textsc{tw} & \textsc{tw+} & \textsc{tp} & \textsc{tp+}
&  \textsc{apc} & \textsc{tw} & \textsc{tw+} & \textsc{tp} & \textsc{tp+} \\ \hline
&\multicolumn{5}{c|}{$(T,N)=(300,500),(T_o,N_o)=(120,300)$} &
\multicolumn{5}{c}{$(T,N)=(500,300),(T_o,N_o)=(300,1200)$} \\ \hline

 &\multicolumn{5}{c}{$C_{115,290}= -1.066 $} &\multicolumn{5}{c}{$C_{190,165}= 1.017 $}\\ \hline
          mean  &   -1.066   &  -1.067   &  -1.066   &  -1.067   &  -1.066   &  1.017   &  1.016   &  1.018   &  1.016   &1.018 \\
             sd &   0.157   &  0.160   &  0.157   &  0.160   &  0.157   &  0.090   &  0.103   &  0.091   &  0.103   &0.090 \\
            ase &   0.145   &  0.155   &  0.132   &  0.154   &  0.145   &  0.079   &  0.123   &  0.077   &  0.119   &0.079 \\
       $q_{05}$ &   -1.769   &  -1.692   &  -1.951   &  -1.698   &  -1.776   &  -1.965   &  -1.505   &  -2.006   &  -1.467   &-1.960 \\
       $q_{95}$ &   1.797   &  1.715   &  1.978   &  1.721   &  1.784   &  1.827   &  1.297   &  1.890   &  1.382   &1.843 \\
       coverage &   0.930   &  0.941   &  0.898   &  0.940   &  0.930   &  0.912   &  0.981   &  0.903   &  0.980   &0.911 \\ \hline

 &\multicolumn{5}{c}{$C_{125,290}= 1.084 $} &\multicolumn{5}{c}{$C_{210,165}= -1.430 $}\\ \hline
          mean  &   1.083   &  1.083   &  1.083   &  1.083   &  1.083   &  -1.429   &  -1.429   &  -1.428   &  -1.429   &-1.428 \\
             sd &   0.101   &  0.106   &  0.104   &  0.106   &  0.104   &  0.100   &  0.112   &  0.107   &  0.112   &0.107 \\
            ase &   0.079   &  0.089   &  0.084   &  0.089   &  0.077   &  0.088   &  0.113   &  0.093   &  0.113   &0.080 \\
       $q_{05}$ &   -2.157   &  -2.070   &  -2.122   &  -2.065   &  -2.306   &  -1.786   &  -1.565   &  -1.816   &  -1.567   &-2.102 \\
       $q_{95}$ &   2.048   &  1.887   &  1.987   &  1.889   &  2.178   &  1.936   &  1.719   &  2.026   &  1.725   &2.337 \\
       coverage &   0.873   &  0.900   &  0.883   &  0.899   &  0.852   &  0.919   &  0.952   &  0.907   &  0.952   &0.858 \\ \hline

 &\multicolumn{5}{c}{$C_{115,325}= -3.260 $} &\multicolumn{5}{c}{$C_{195,195}= 0.923 $}\\ \hline
       mean  &   -3.255   &  -3.245   &  -3.249   &  -3.242   &  -3.252   &  0.922   &  0.917   &  0.920   &  0.917   &0.921 \\
             sd &   0.174   &  0.257   &  0.212   &  0.216   &  0.207   &  0.092   &  0.129   &  0.106   &  0.115   &0.104 \\
            ase &   0.162   &  0.289   &  0.175   &  0.220   &  0.190   &  0.080   &  0.107   &  0.089   &  0.125   &0.085 \\
       $q_{05}$ &   -1.750   &  -1.407   &  -1.960   &  -1.525   &  -1.789   &  -1.953   &  -2.108   &  -2.070   &  -1.601   &-2.109 \\
       $q_{95}$ &   1.832   &  1.538   &  2.066   &  1.718   &  1.828   &  1.843   &  1.826   &  1.835   &  1.414   &1.889 \\
       coverage &   0.932   &  0.972   &  0.889   &  0.951   &  0.922   &  0.910   &  0.895   &  0.899   &  0.967   &0.895 \\ \hline

 &\multicolumn{5}{c}{$C_{140,325}= 2.451 $} &\multicolumn{5}{c}{$C_{220,205}= -1.115 $}\\ \hline

       mean  &   2.448   &  2.442   &  2.445   &  2.440   &  2.444   &  -1.120   &  -1.110   &  -1.115   &  -1.112   &-1.116 \\
             sd &   0.145   &  0.210   &  0.187   &  0.188   &  0.183   &  0.138   &  0.214   &  0.184   &  0.179   &0.177 \\
            ase &   0.129   &  0.225   &  0.153   &  0.173   &  0.135   &  0.127   &  0.180   &  0.133   &  0.158   &0.148 \\
       $q_{05}$ &   -1.874   &  -1.653   &  -2.120   &  -1.911   &  -2.349   &  -1.847   &  -1.884   &  -2.209   &  -1.841   &-1.984 \\
       $q_{95}$ &   1.837   &  1.452   &  1.951   &  1.692   &  2.181   &  1.772   &  2.032   &  2.322   &  1.901   &1.971 \\
       coverage &   0.916   &  0.963   &  0.885   &  0.922   &  0.847   &  0.922   &  0.899   &  0.847   &  0.915   &0.897 \\ \hline

\end{tabular}
\end{center}
Data are generated from a model with two factors. The factors and the loadings are drawn from the normal distribution once and hence fixed. In each of the 5000 replications, new idiosyncratic errors are resampled. The location of the missing values are fixed throughout.  The \textsc{apc} column is the infeasible case when the data are completely observed. The columns \textsc{tw} and \textsc{tp} are one step estimators. A '+' indicates one re-estimation.
\end{table}


\begin{table}
	\caption{Monte Carlo:  Strict Factor Model, $15\%$ Missingness,  $T=339, N=100$}
	\label{tbl:simdata2a}
	\begin{center}
		\begin{tabular}{l|ccccc|ccccc} \hline
			
			& \multicolumn{5}{c|}{Bias}
			&  \multicolumn{5}{|c}{RMSE}  \\ \hline	
			
			\hline
			& pvol   &   pVaR   &   call  & var  & covar
			& pvol   &   pVaR   &   call  & var  & covar
			\\ \hline
 & \multicolumn{10}{c}{TW} \\ \hline			
sm0  &-0.037 &-0.078 &-1.105 &-0.167 & 0.000 & 0.041 & 0.088 & 1.231 & 0.212 & 0.031 \\
     sm1  &-0.003 &-0.011 & 0.068 &-0.019 & 0.000 & 0.017 & 0.040 & 0.815 & 0.131 & 0.031 \\
     sm2  & 0.003 &-0.001 & 0.038 & 0.005 & 0.000 & 0.016 & 0.037 & 0.388 & 0.072 & 0.031 \\
     sm3  &-0.003 &-0.011 & 0.068 &-0.019 & 0.000 & 0.017 & 0.040 & 0.815 & 0.131 & 0.031 \\
     sm4  & 0.003 &-0.000 & 0.044 & 0.006 & 0.000 & 0.016 & 0.037 & 0.388 & 0.072 & 0.031 \\
     sfa  & 0.003 &-0.000 & 0.100 & 0.015 & 0.000 & 0.023 & 0.049 & 0.535 & 0.099 & 0.042 \\
    sm+0  &-0.036 &-0.074 &-1.047 &-0.157 & 0.000 & 0.040 & 0.082 & 1.152 & 0.195 & 0.029 \\
    sm+1  &-0.003 &-0.009 & 0.080 &-0.015 & 0.000 & 0.016 & 0.034 & 0.748 & 0.116 & 0.029 \\
    sm+2  &-0.003 &-0.008 &-0.070 &-0.013 & 0.000 & 0.015 & 0.033 & 0.369 & 0.069 & 0.029 \\
    sm+3  &-0.003 &-0.009 & 0.080 &-0.015 & 0.000 & 0.016 & 0.034 & 0.748 & 0.116 & 0.029 \\
    sm+4  &-0.003 &-0.008 &-0.066 &-0.012 & 0.000 & 0.015 & 0.033 & 0.368 & 0.069 & 0.029 \\
    sf+a  &-0.009 &-0.020 &-0.135 &-0.025 &-0.000 & 0.019 & 0.041 & 0.427 & 0.083 & 0.030 \\

\hline
 & \multicolumn{10}{c}{TP} \\ \hline			
\hline

     sm0  &-0.038 &-0.079 &-1.129 &-0.170 & 0.000 & 0.042 & 0.088 & 1.239 & 0.212 & 0.028 \\
     sm1  &-0.004 &-0.013 & 0.040 &-0.024 & 0.000 & 0.016 & 0.037 & 0.795 & 0.128 & 0.028 \\
     sm2  & 0.001 &-0.002 & 0.003 & 0.000 & 0.000 & 0.015 & 0.035 & 0.349 & 0.064 & 0.028 \\
     sm3  &-0.004 &-0.013 & 0.040 &-0.024 & 0.000 & 0.016 & 0.037 & 0.795 & 0.128 & 0.028 \\
     sm4  & 0.001 &-0.002 & 0.008 & 0.001 & 0.000 & 0.015 & 0.035 & 0.348 & 0.064 & 0.028 \\
     sfa  &-0.001 &-0.007 & 0.011 & 0.001 & 0.000 & 0.017 & 0.039 & 0.348 & 0.064 & 0.029 \\
    sm+0  &-0.036 &-0.074 &-1.053 &-0.158 & 0.000 & 0.040 & 0.082 & 1.152 & 0.194 & 0.028 \\
    sm+1  &-0.004 &-0.010 & 0.074 &-0.016 & 0.000 & 0.015 & 0.033 & 0.741 & 0.114 & 0.028 \\
    sm+2  &-0.003 &-0.009 &-0.077 &-0.014 & 0.000 & 0.015 & 0.033 & 0.359 & 0.067 & 0.028 \\
    sm+3  &-0.004 &-0.010 & 0.074 &-0.016 & 0.000 & 0.015 & 0.033 & 0.740 & 0.114 & 0.028 \\
    sm+4  &-0.003 &-0.009 &-0.073 &-0.013 & 0.000 & 0.015 & 0.033 & 0.358 & 0.067 & 0.028 \\
    sf+a  &-0.010 &-0.022 &-0.152 &-0.027 &-0.000 & 0.018 & 0.040 & 0.398 & 0.077 & 0.028 \\

\hline\hline
\end{tabular}
\end{center}

\end{table}

\begin{table}
	\caption{Additional TP Results:  Strict Factor Model}
	\label{tbl:simdata2b}
	\begin{center}
		\begin{tabular}{l|ccccc|ccccc} \hline
			
			& \multicolumn{5}{c|}{Bias}
			&  \multicolumn{5}{|c}{RMSE}  \\ \hline	
			
			\hline
			& pvol   &   pVaR   &   call  & var  & covar
			& pvol   &   pVaR   &   call  & var  & covar

			\\ \hline

 & \multicolumn{10}{c}{30\% Missing, $(T,N)=(339,100)$} \\ \hline			
\hline
    sm0  &-0.082 &-0.163 &-1.601 &-0.236 & 0.000 & 0.086 & 0.174 & 1.739 & 0.289 & 0.040 \\
    sm1  &-0.010 &-0.023 & 0.003 &-0.037 & 0.000 & 0.026 & 0.060 & 1.045 & 0.170 & 0.040 \\
    sm2  & 0.003 & 0.003 & 0.003 & 0.001 & 0.000 & 0.023 & 0.055 & 0.447 & 0.081 & 0.040 \\
    sm3  &-0.010 &-0.023 & 0.003 &-0.037 & 0.000 & 0.026 & 0.060 & 1.045 & 0.170 & 0.040 \\
    sm4  & 0.003 & 0.003 & 0.011 & 0.002 & 0.000 & 0.023 & 0.055 & 0.447 & 0.081 & 0.040 \\
    sfa  & 0.004 & 0.005 & 0.016 & 0.003 & 0.000 & 0.026 & 0.060 & 0.447 & 0.081 & 0.042 \\
   sm+0  &-0.080 &-0.157 &-1.516 &-0.223 & 0.000 & 0.084 & 0.167 & 1.642 & 0.270 & 0.038 \\
   sm+1  &-0.013 &-0.026 &-0.015 &-0.036 &-0.000 & 0.026 & 0.054 & 0.970 & 0.157 & 0.038 \\
   sm+2  &-0.010 &-0.020 &-0.167 &-0.028 & 0.000 & 0.024 & 0.050 & 0.478 & 0.088 & 0.038 \\
   sm+3  &-0.013 &-0.026 &-0.015 &-0.036 &-0.000 & 0.026 & 0.054 & 0.970 & 0.157 & 0.038 \\
   sm+4  &-0.009 &-0.019 &-0.159 &-0.027 & 0.000 & 0.024 & 0.050 & 0.475 & 0.088 & 0.038 \\
   sf+a  &-0.016 &-0.032 &-0.259 &-0.044 & 0.000 & 0.028 & 0.058 & 0.536 & 0.101 & 0.039 \\
\hline
 & \multicolumn{10}{c}{$15\%$ Missing,  $(T,N)=(200, 250)$} \\ \hline			
     sm0  &-0.015 &-0.034 &-0.607 &-0.092 & 0.000 & 0.018 & 0.041 & 0.906 & 0.153 & 0.029 \\
     sm1  &-0.001 &-0.006 & 0.092 &-0.005 & 0.000 & 0.009 & 0.022 & 0.780 & 0.122 & 0.029 \\
     sm2  & 0.000 &-0.005 &-0.007 &-0.001 & 0.000 & 0.009 & 0.021 & 0.349 & 0.061 & 0.029 \\
     sm3  &-0.001 &-0.006 & 0.093 &-0.005 & 0.000 & 0.009 & 0.022 & 0.780 & 0.122 & 0.029 \\
     sm4  & 0.000 &-0.005 &-0.002 &-0.001 & 0.000 & 0.009 & 0.021 & 0.349 & 0.061 & 0.029 \\
     sfa  &-0.001 &-0.007 & 0.001 &-0.000 & 0.000 & 0.009 & 0.020 & 0.349 & 0.061 & 0.028 \\
    sm+0  &-0.015 &-0.032 &-0.589 &-0.089 &-0.000 & 0.018 & 0.038 & 0.881 & 0.148 & 0.029 \\
    sm+1  &-0.001 &-0.005 & 0.102 &-0.003 &-0.000 & 0.009 & 0.020 & 0.762 & 0.118 & 0.029 \\
    sm+2  &-0.001 &-0.005 &-0.023 &-0.004 &-0.000 & 0.009 & 0.020 & 0.349 & 0.061 & 0.029 \\
    sm+3  &-0.001 &-0.005 & 0.102 &-0.003 &-0.000 & 0.009 & 0.020 & 0.762 & 0.118 & 0.029 \\
    sm+4  &-0.001 &-0.005 &-0.018 &-0.003 &-0.000 & 0.009 & 0.020 & 0.349 & 0.061 & 0.029 \\
    sm+d  &-0.002 &-0.008 &-0.033 &-0.006 &-0.000 & 0.009 & 0.019 & 0.352 & 0.062 & 0.028 \\
\hline
\end{tabular}

\end{center}
\end{table}


\begin{table}
	\caption{Monte Carlo Calibrated to CRSP Data: $15\%$ Missing }
	\label{tbl:sp500res2}
	\begin{center}
		\begin{tabular}{l|ccccc|ccccc} \hline
			
			& \multicolumn{5}{c|}{Bias}
			&  \multicolumn{5}{|c}{RMSE}  \\ \hline	
			
			\hline
			& pvol   &   pVaR   &   call  & var  & covar
			& pvol   &   pVaR   &   call  & var  & covar
			\\ \hline
 & \multicolumn{10}{c}{ TW  } \\ \hline			
    sm0  &-0.006 &-0.025 &-1.399 &-0.211 & 0.001 & 0.074 & 0.147 & 2.131 & 0.396 & 0.068 \\
     sm1  & 0.004 &-0.006 & 0.392 & 0.012 & 0.001 & 0.073 & 0.145 & 1.794 & 0.335 & 0.068 \\
     sm2  & 0.007 &-0.000 & 0.532 & 0.080 & 0.001 & 0.073 & 0.144 & 1.647 & 0.333 & 0.068 \\
     sm3  & 0.004 &-0.006 & 0.392 & 0.012 & 0.001 & 0.073 & 0.145 & 1.794 & 0.335 & 0.068 \\
     sm4  & 0.007 &-0.000 & 0.540 & 0.081 & 0.001 & 0.073 & 0.144 & 1.650 & 0.334 & 0.068 \\
     sfa  & 0.029 & 0.043 & 0.788 & 0.122 &-0.000 & 0.116 & 0.225 & 2.049 & 0.410 & 0.106 \\
     sm+  & 0.010 & 0.026 &-1.393 &-0.206 & 0.003 & 0.059 & 0.117 & 2.050 & 0.378 & 0.062 \\
    sm+1  & 0.018 & 0.043 & 0.332 & 0.008 & 0.003 & 0.061 & 0.122 & 1.673 & 0.317 & 0.062 \\
    sm+2  & 0.019 & 0.045 & 0.266 & 0.032 & 0.003 & 0.061 & 0.122 & 1.438 & 0.289 & 0.062 \\
    sm+3  & 0.018 & 0.043 & 0.333 & 0.008 & 0.003 & 0.061 & 0.122 & 1.673 & 0.317 & 0.062 \\
    sm+4  & 0.019 & 0.045 & 0.273 & 0.033 & 0.003 & 0.061 & 0.122 & 1.440 & 0.289 & 0.062 \\
    sfa+  & 0.065 & 0.135 & 0.237 & 0.022 & 0.006 & 0.098 & 0.197 & 1.540 & 0.304 & 0.082 \\

\hline
 & \multicolumn{10}{c}{ TP} \\ \hline			
\hline
      sm0  &-0.019 &-0.047 &-1.553 &-0.229 &-0.001 & 0.049 & 0.100 & 2.076 & 0.380 & 0.052 \\
     sm1  &-0.010 &-0.028 & 0.231 &-0.011 &-0.001 & 0.046 & 0.093 & 1.582 & 0.303 & 0.052 \\
     sm2  &-0.007 &-0.023 & 0.323 & 0.048 &-0.001 & 0.046 & 0.091 & 1.381 & 0.286 & 0.052 \\
     sm3  &-0.010 &-0.028 & 0.231 &-0.011 &-0.001 & 0.046 & 0.093 & 1.582 & 0.303 & 0.052 \\
     sm4  &-0.007 &-0.023 & 0.331 & 0.049 &-0.001 & 0.046 & 0.091 & 1.384 & 0.287 & 0.052 \\
     sfa  & 0.004 &-0.001 & 0.336 & 0.050 &-0.003 & 0.045 & 0.088 & 1.385 & 0.287 & 0.068 \\
     sm+  & 0.003 & 0.015 &-1.458 &-0.213 & 0.002 & 0.044 & 0.088 & 1.998 & 0.363 & 0.053 \\
    sm+1  & 0.011 & 0.031 & 0.273 &-0.001 & 0.002 & 0.046 & 0.092 & 1.551 & 0.294 & 0.053 \\
    sm+2  & 0.012 & 0.033 & 0.198 & 0.022 & 0.002 & 0.046 & 0.092 & 1.315 & 0.266 & 0.053 \\
    sm+3  & 0.011 & 0.031 & 0.273 &-0.001 & 0.002 & 0.046 & 0.092 & 1.551 & 0.294 & 0.053 \\
    sm+4  & 0.012 & 0.033 & 0.205 & 0.023 & 0.002 & 0.046 & 0.092 & 1.317 & 0.266 & 0.053 \\
    sfa+  & 0.052 & 0.111 & 0.115 & 0.005 & 0.004 & 0.069 & 0.142 & 1.329 & 0.269 & 0.068 \\

			\hline\hline

\end{tabular}
\end{center}

\end{table}

\clearpage
\baselineskip=12.0pt
\bibliography{factor,metrics,macro,forecast,metrics2,bigdata,macro1}

\baselineskip=15.0pt
\newpage
\section*{Appendix}
\newcommand{\lambdaUpperCase}{\Lambda}  
\newcommand{\tL}{\tilde \lambdaUpperCase}
\newcommand{\hF}{\hat F}
\newcommand{\hL}{\hat \lambdaUpperCase^+}
\newcommand{\breveL}{\tilde \lambdaUpperCase^+}
\renewcommand{\breve}{\tilde}

\newcommand{\calE}{{\cal E}}
\newcommand{\mathD}{{\mathbb D}}
\newcommand{\mathQ}{\mathbb Q}
\newcommand{\Lambdaop}{\Lambda^{0\prime}}   
\newcommand{\Fop}{ F^{0\prime}}
\renewcommand{\To}{T_o}  

\newcommand{\Lambdao}{\Lambda_o}   
\newcommand{\Lambdam}{\Lambda_m}    
\newcommand{\Fo}{F_o}
\newcommand{\Fm}{F_m}
\newcommand{\tF}{\tilde F}
\newcommand{\hFplusp}{\hat F^{+\prime}}
\newcommand{\hFplus}{\hat F^+}
\newcommand{\oh}{{o_h}}
\newcommand{\breveFplusp}{\tilde F^{+\prime}}

 We observe $X$, but not $F^0$ or $\Lambda^0$. As $F$ and $\Lambda$ are not separately identifiable. The method of asymptotic principal components (APC) uses the normalization  $\frac{F\pr F}{T}=I_r$ and
 $\Lambda'\Lambda$ being diagonal to produce  estimates
 \[ (\tilde F,\tilde \Lambda)=(\sqrt{T} U_r,\sqrt{N}V_rD_r). \]
 For each $t\in [1,T]$ and for each $i\in[1,N]$,    $(\tilde F_t$, $\tilde \Lambda_i)$ consistently estimate $(F_t^0$, $\Lambda^0_i$)   up to a rotation matrices $H$. This rotation matrix is not unqiue, and for the present analysis, let
  \begin{eqnarray*}    H&=&\bigg(\frac{\Lambda^{0^\prime}\Lambda^0}{N}\bigg)\bigg(\frac{F^{0^\prime}\tilde F}{T}\bigg) D_r^{-2}
    \end{eqnarray*}
denote the rotation matrix for  complete data, and let
 \begin{eqnarray*}    H_{tall}&=&\bigg(\frac{\Lambda_o^{0^\prime}\Lambda_o^0}{N_o}\bigg)\bigg(\frac{F^{0^\prime}\tilde F_{tall}}{T}\bigg) D_{r,tall}^{-2}
    \end{eqnarray*}
 denote the rotation matrix for the tall block of data, where $\Lambda_o^0$ is the $N_o \times r$ matrix consisting of factor loadings of $\Lambda_k^0$ for all $k\in \cap_s J^s$,
 and $\tilde F_{tall}$ is $T\times r$, and $D_{r,tall}^2$ is $r\times r$.

\paragraph{Proof of Lemma \ref{lem:tp}.}
By construction, the first stage estimated factors are obtained
 from the observed $T\times \No $ data matrix, giving rise to
$\tilde F_{\text{tall}}$.
For $(i,t)$ in the tall block, the regression method and the principal components estimator  are the same as the complete data case  explained in the text. So Lemma \ref{lem:tp}i, and ii(a) follow from Lemma 1. We consider the remaining claims.

For each $i>N_o$ of the reorganized data (or  $i\not \in \cap_s J^s$ of the  original data),  we observe $T_{o_i}$ observations. Here for notational simplicity, we assume the first consecutive $T_{o_i}$ observations are available. The true model is
$X_{it}=\Fop_t\Lambda_i^0+e_{it}$. To estimate the factor loadings $\Lambda_i^0$, consider the regression
\[ X_{it} =\tF_t'\Lambda_i + a_{it}, \quad t=1,2,...,T_{o_i}  \]
where $\tilde F_t'$ is the $t$th row of $\tilde F_{\text{tall}}$,  $a_{it}$ is an error term.    Then, OLS gives
\[ \breve{\Lambda}_i= (\tF_\oi'\tF_\oi)^{-1}\tF_\oi'X_\oi \]
where $X_\oi$ is $T_\oi\times 1$ vector such that $X_\oi=F_\oi^0\Lambda_i^0+e_{\oi}$, and
$F_\oi^0$ is $T_\oi \times r$, and $e_\oi$ is $T_\oi\times 1$;  $\tF_\oi$ stacks $\tilde F_t$.
It follows that
\[ \breve \Lambda_i =(\tF_\oi'\tF_\oi)^{-1}\tF_\oi'F_\oi^0\Lambda_i^0 + (\tF_\oi'\tF_\oi)^{-1}\tF_\oi'e_\oi. \]
Let $G_i=(\tF_\oi'\tF_\oi)^{-1}\tF_\oi'F_\oi^0$ and  rewrite
$F_\oi=\tF_\oi H_{tall}^{-1}-[\tF_\oi-F_\oi H_{tall} ] H_{tall}^{-1}$, so that
\[G_i=H_{tall}^{-1}-(\tF_\oi'\tF_\oi)^{-1}\tF_\oi'[\tF_\oi-F_\oi H_{tall} ] H_{tall}^{-1}.\] Since
$(\tF_\oi'\tF_\oi/T_\oi)^{-1}=O_p(1)$, and $\frac 1 {T_\oi} \tF_\oi'[\tF_\oi-F_\oi H_{tall} ]=O_p(\delta_{N_o,T_\oi}^{-2})$, we obtain  \[ G_i=H_{tall}^{-1}+O_p(\delta_{N_o,T_\oi}^{-2}).\]   Thus,
\begin{align} \breve \Lambda_i -H_{tall}^{-1} \Lambda_i^0 & = (\tF_\oi'\tF_\oi)^{-1}\tF_\oi'e_\oi + O_p(1/\delta_{N_o T_{o_i}}^2) \nonumber \\
 & =  H_{tall}^{-1} (F_\oi^{0\pr}F_\oi^0)^{-1} F_\oi^{0\pr}e_\oi +O_p(1/\delta_{N_o T_{o_i}}^2).
\end{align}
Multiply by $\sqrt{T_{o_i}}$ and use the limit for $H_{tall}$ (see \citet{baing-miss:jasa}), we obtain part ii(b) of Lemma \ref{lem:tp}. Here we use the assumption that  $\frac{\sqrt{T_{\oi}}}{N_o}\rightarrow 0$.

Consider the estimated common components
\[ \breve C_{it} =\tF_t'\breve \Lambda_i
 =\tF_t' (\tF_\oi'\tF_\oi)^{-1}\tF_\oi'F_\oi^0\Lambda_i^0+\tF_t'(\tF_\oi'\tF_\oi)^{-1}\tF_\oi'e_\oi . \]
Similar to \citet{baing-miss:jasa}, we can show, for all missing entries $(i,t)$,
\begin{eqnarray}  \breve C_{it}&=& C_{it}^0 + \lambdaUpperCase_i^{0\prime}(\Lambda_o^{0\prime}\Lambda_o^0/\No)^{-1} \frac 1 {N_o}\sum_{k=1}^{\No} \Lambda_k^0 e_{kt} + F_t^{0\prime}(F_\oi^{0\prime}F_\oi^0/T_\oi)^{-1} \frac 1 {T_\oi}\sum_{s=1}^{T_\oi} F_s^0 e_{is} + O_p(1/\delta_{N_o,T_\oi}^2).\nonumber
 \end{eqnarray}
 where $\Lambda_o^0$ is $N_o\times r$. Here and for the remainder of the proof, we assume the first $N_o$ units have no missing observations so that $\sum_{k=1}^{N_o}$ is identical to
 $\sum_{k\in \cap_s J^s}$. Let
 \begin{align} u_{it} & =  \lambdaUpperCase_i^{0\prime}(\Lambda_o^{0\prime}\Lambda_o^0/\No)^{-1} \frac 1 {N_o}\sum_{k=1}^{\No} \Lambda_k^0 e_{kt} \\ \label{eq:vit}
  v_{it} & =  F_t^{0\prime}(F_\oi^{0\prime}F_\oi^0/T_\oi)^{-1} \frac 1 {T_\oi}\sum_{s=1}^{T_\oi} F_s^0 e_{is}
 \end{align}
 Then
  \begin{eqnarray}
\breve C_{it}  &=& C_{it}^0+u_{it} +  v_{it} +O_p(1/\delta_{N_o,T_\oi}^2).
 \label{breveCit1}
  \end{eqnarray}
 This implies that
\begin{equation}\label{breveC-C0-initial}  \Big(\frac 1 {N_o}  A_{NT} +\frac 1 {T_\oi} B_{NT}\Big)^{-1/2}
(\breve C_{it}-C_{it}^0 )\dconv
 N(0,1) \end{equation}
where
$ A_{NT}= \lambdaUpperCase_i^{0\prime}(\Lambda_o^{0\prime}\Lambda_o^0/\No)^{-1}\Gamma_t
(\Lambda_o^{0\prime}\Lambda_o^0/\No)^{-1} \lambdaUpperCase_i^0$ and
$ B_{NT}= F_t^{0\prime}(F_\oi^{0\prime}F_\oi^0/T_\oi)^{-1}\Phi_i(F_\oi^{0\prime}F_\oi^0/T_\oi)^{-1}  F_t^0$.
This gives  part iii of Lemma \ref{lem:tp}.
\subsection{Proof of Proposition 1}

Consider the principal components estimator of factors and factor loadings based on $\breve X$ (see equations
(\ref{eq:Xtilde-obs}) and (\ref{eq:Xtilde-miss})). Let
$\breve F^+$ and $\breve \Lambda^+$ denote the PCA estimator so that
\[ \frac 1 {NT} \breve X \breve X' \breve F^+ = \breve F^+ \breve D_r^2 \]
with $\breve F^{+\prime} \breve F^+/T=I_r$,
and $\breve \Lambda = \frac 1 T \breve X'\breve F^+$; $\breve D_r^2$ is a diagonal matrix consisting of the first
$r$ largest eigenvalues of $\breve X \breve X'/(NT)$.  Define
\[  H^+  = (\Lambda^{0\prime}\Lambda^0/N)(F^{0\prime} \breve F/T) \breve D_r^{-2}  \]


Now we assume $T_\oi\ge T_o$ for some $T_o$. We also assume $N_{o_t}\ge N_o$, where $T_o$ and $N_o$ satisfy Assumption B.
Then Lemma 3 in \citet{baing-miss:jasa}  holds.  That is

\begin{lemma} \label{TPlemma1} Suppose that $N_{o_t}\ge N_o$ and $T_\oi\ge T_o$ for all $t$ and $i$. Under Assumptions A and B, we have
(a) $\frac 1 T \sum_{t=1}^T \| \breve F_t^+- H^+  F_t^0\|^2 =O_p( \delta_{N_o,T_o}^{-2}) $, and
(b) $\plim \tilde D_r^2\rightarrow \mathD_r^2$ and $  \breve F^{+\prime}F^0/T \rightarrow \mathbb Q_r$, where
$\mathD_r$ and $\mathbb Q$ are the same as in complete data.
\end{lemma}

The proof of this lemma follows the same argument as in \citet{baing-miss:jasa}. The details are omitted. We focus on obtaining the asymptotic distributions.

{\bf Asymptotic distribution for the estimated factors}.
To derive the limiting distribution for the estimated factor and factor loadings,
 define
\[ \breve e_{it}= \left\{ \begin{array}{ll} e_{it} & \text{ if } X_{it} \text{ is observed } \\
u_{it}+ v_{it}  & \text{ if } X_{it} \text{ is missing } \end{array}  \right. \]
Because of the preceding lemma, the asymptotic representation for $\breve F^+_t$ is (see Bai (2003) and \citet{baing-miss:jasa}),
\begin{equation} \label{breveFplus-rep}
 \breve F^+_t - H^{+\prime} F_t^0=
\breve D^{-2} (\breveFplusp F^0/T) \frac 1 N \sum_{i=1}^N \lambdaUpperCase_i^0 \breve e_{it}  +O_p(1/\delta_{\No ,\To}^2). \end{equation}
For a given $t$, if all individual variables $(X_{1t}, X_{2t},...,X_{Nt})$ are observable, then
$\breve e_{it}=e_{it}$ for all $i$, and
\begin{equation} \label{breveF-full-N-limit} \sqrt{N} ( \breve F^+_t - H^{+\prime} F_t^0)=\breve D^{-2} (\breveFplusp F^0/T) \frac 1 {\sqrt{N}} \sum_{i=1}^N \lambdaUpperCase_i^0 e_{it}  +O_p(1/\delta_{\No ,\To}^2) \end{equation}
\[ \dconv N(0, \mathD_r^{-2} \mathQ_r \Gamma_t \mathQ_r'\mathD_r^{-2}). \]
For a given $t$,  suppose that (the first) $N_{o_t}$ series are available, denoted by
$(X_{1t},X_{2t}, ... X_{N_{o_t} t})$,  with $N_{o_t} \ge N_o$. Then $\breve e_{it}=e_{it}$ for $i\le N_{o_t}$ and
$\breve e_{it}=u_{it}+ v_{it}$ for $i> N_{o_t}$, so (\ref{breveFplus-rep}) becomes
\[ \breve F^+_t - H^{+\prime} F_t^0 =\breve D^{-2} (\breveFplusp F^0/T)  \Big( \frac 1 N \sum_{i=1}^{N_{o_t}} \lambdaUpperCase_i^0 e_{it} + \frac 1 N \sum_{i=N_{o_t}+1}^N \lambdaUpperCase_i^0(u_{it}+ v_{it})\Big)
 +O_p(1/\delta_{\No ,\To}^2). \]
 Note that  $\frac 1 N \sum_{i=N_{o_t}+1}^N \lambdaUpperCase_i^0 v_{it}$ is negligible  (dominated by $\frac 1 N \sum_{i=N_{o_t}+1}^N \lambdaUpperCase_i^0 u_{it}$). This follows from the definition of $v_{it}$ in
  (\ref{eq:vit}),
 \[ \frac 1 N \sum_{i=N_t+1}^N \Lambda_{i,j}^0 v_{it} = O_p(1)  \Big( \frac 1 {N T_{o_i} } \sum_{i=N_t+1}^N \sum_{s=1}^{T_\oi} \Lambda_{i,j}^0 F_s^0 e_{is}\Big)
    =O_p(1/\sqrt{N T_{o_i}})=O_p(1/\delta_{N_o,T_o}^2) \]
    where $\Lambda_{i,j}^0$ is the jth component of $\Lambda_i^0$ ($j=1,2,...,r$).

 Let
 \[ A_t =I_r+ \Big(\frac{N-N_{o_t}}{\No} \Big)\Big(\frac 1 {N-N_{o_t}}\sum_{i=N_{o_t}+1}^N \lambdaUpperCase_i^0\lambdaUpperCase_i^{0\prime}\Big) (\Lambda_o^{0\prime}\Lambda_o^0/\No)^{-1}.  \]
 We can rewrite the representation as
 \[ \breve F^+_t - H^{+\prime} F_t^0=\breve D^{-2} (\breveFplusp F^0/T)  \Big( \frac 1 N \sum_{i=1}^{N_o}
 A_t \lambdaUpperCase_i^0 e_{it} + \frac 1 N \sum_{i=N_o+1}^{N_{o_t}} \lambdaUpperCase_i^0 e_{it}\Big)  +O_p(1/\delta_{\No ,\To}^2), \]
For the special case that $N_{o_t}=N_o$, and $\Big(\frac 1 {N-N_{o_t}}\sum_{i=N_{o_t}+1}^N \lambdaUpperCase_i^0\lambdaUpperCase_i^{0\prime} \Big)\Big(\Lambda_o^{0\prime}\Lambda_o^0/\No\Big)^{-1} \rightarrow I_r$
 we have $A_t \sim N/N_o$, the above representation coincides with the TW presentation in
\citet{baing-miss:jasa}.

Consider the more general case of  $N_{o_t}$ with $N_{o_t} \ge N_o$. Define
 \[  B_t^i= \left\{ \begin{array}{ll} \frac {N_{o_t}} N  A_t & i \le N_o  \vspace{0.1in} \\
 \frac {N_{o_t}} N I_r & N_o<i \le N_{o_t} \end{array} \right. \]
 We can further rewrite the representation as
 \begin{equation}\label{F+Rep}\breve F^+_t - H^{+\prime} F_t^0 =\breve D^{-2} (\breveFplusp F^0/T)  \Big( \frac 1 {N_{o_t}}  \sum_{i=1}^{N_{o_t}} B_t^i \lambdaUpperCase_i^0 e_{it} \Big)
 +O_p(1/\delta_{\No ,\To}^2). \end{equation}
 Under Assumption C,  $B_t^i$ is bounded. Then the convergence rate is
$\sqrt{N_{o_t}}$, and
\[ \sqrt{N_{o_t}} ( \breve F^+_t - H^{+\prime} F_t^0)=\breve D^{-2} (\breveFplusp F^0/T)\frac 1 {\sqrt{N_{o_t}}} \sum_{i=1}^{N_{o_t}} B_t^i \lambdaUpperCase_i^0 e_{it}  +o_p(1) \]
\[ \dconv  N(0,  \mathD_r^{-2} \mathQ_r \Gamma_t^* \mathQ_r'\mathD_r^{-2}), \]
where
\begin{equation} \label{Gammat*} \Gamma_t^*= \plim \frac 1 {N_{o_t}} \sum_{i=1}^{N_{o_t}} B_t^i \lambdaUpperCase_i^0\lambdaUpperCase_i^{0\prime}  B_t^{i\prime} e_{it}^2 \end{equation}
(assuming cross-sectional uncorrelation for $e_{it}$).

The result includes  $N_{o_t}=N$ as a special case. When $N_{o_t}=N$, we have $B_t^i=I_r$ for all $i$, and
$\Gamma_t^*=\Gamma_t$.  The convergence is
$\sqrt{N}$, and the limit is in (\ref{breveF-full-N-limit}).


\paragraph{Asymptotic distribution for the estimated factor loadings:}
We have the asymptotic representation
\[  \breveL_i- (H^{+})^{-1} \lambdaUpperCase_i^0=  H^{+\prime}  \frac 1 T \sum_{t=1}^T F_t^0 \breve e_{it} +\breve \eta_{NT,i}  \]
where $ \breve \eta_{NT,i} =O_p(\delta_{N_o,T_o}^{-2})$ is uniformly in $i$ and $t$.

For a given $i$, if all  $T$ observations are available, then $\breve e_{it}=e_{it}$ for all $t$, so that
\[  \sqrt{T} ( \breveL_i- (H^{+})^{-1} \lambdaUpperCase_i^0)=  H^{+\prime}  \frac 1 {\sqrt{T}} \sum_{t=1}^T F_t^0 e_{it} + o_p(1)\rightarrow^d N(0, \mathQ_r^{\prime -1} \Phi_i \mathQ_r^{-1})   \]
where we used the fact that the limit of $H^+$ is $\mathQ_r^{-1}$.

Suppose that for the given $i$, there are $T_\oi$ observations available.  Again, for notational simplicity, we assume the first
$T_{o_i}$ are observable, the rest are missing.
Then $\breve e_{it}=e_{it}$ for $t\le T_\oi$ and $\breve e_{it}=u_{it}+ v_{it}$ for $t >T_\oi$. Thus
\[ \breveL_i- (H^{+})^{-1} \lambdaUpperCase_i^0=  H^{+\prime} \Big( \frac 1 T \sum_{t=1}^{T_\oi} F_t^0  e_{it}
+\frac 1 T \sum_{t=T_\oi+1}^T F_t^0(u_{it}+ v_{it})\Big)  +\breve \eta_{NT,i} \]
Note that $\frac 1 T \sum_{t=T_\oi+1}^T F_t^0u_{it}$ is negligible, and
\[ \frac 1 T \sum_{t=T_\oi+1}^T F_t^0  v_{it}= \frac {T-T_\oi} T \Big(\frac 1 {T-T_\oi}  \sum_{t=T_\oi+1}^T F_t^0 F_t^{0\prime}\Big) \Big(F_\oi^{0\prime}F_\oi^0/T_\oi\Big)^{-1} \frac 1 {T_\oi}\sum_{s=1}^{T_\oi} F_s^0 e_{is}. \]
Under
\[ \Big(\frac 1 {T-T_\oi}  \sum_{t=T_\oi+1}^T F_t^0 F_t^{0\prime}\Big) \Big(F_\oi^{0\prime}F_\oi^0/T_\oi\Big)^{-1}
\pconv I_r \]
 the asymptotic representation can be written as
\begin{equation}\label{Lambda+Rep}  \breveL_i- (H^{+})^{-1} \lambdaUpperCase_i^0=  H^{+\prime}
 \Big( \frac 1 {T_\oi} \sum_{t=1}^{T_\oi} F_t^0  e_{it}
\Big) + O_p(\delta_{N_o,T_o}^{-2}).  \end{equation}
Then by Assumption D,
\[ \sqrt{T_\oi}(\breveL_i- (H^{+})^{-1} \lambdaUpperCase_i^0) \rightarrow^d N(0, \mathQ_r'\Phi_i \mathQ_r). \]

\paragraph{Asymptotic distribution for the estimated common components:} Let
$ \breve C_{it}^+= \breve F_t^{+\prime} \breveL_i. $
Using the asymptotic representations in (\ref{F+Rep}) and (\ref{Lambda+Rep}), we can show as in \citet{baing-miss:jasa}

\[ \tilde C_{it}^+-C_{it}^0= \Lambda_i^{0\prime}  (\Lambda^{0\prime}\Lambda^0/N) ^{-1}   \Big( \frac 1 {N_{o_t}} \sum_{i=1}^{N_{o_t}} B_t^i \lambdaUpperCase_i^0 e_{it}\Big)
+ F_t^{0\prime} (F^{0\prime}F^0/T)^{-1} \Big( \frac 1 {T_\oi} \sum_{t=1}^{T_\oi} F_t^0  e_{it}
\Big) + O_p(\delta_{N_o,T_o}^{-2}) . \]
The above representation for $\tilde C_{it}^+-C_{it}^0$ implies
 \begin{equation} \label{breve-Cit-Nt} \Big( \frac 1 {N_{o_t}}  V_{it}^*  +\frac 1 {T_\oi}  W_{it} \Big)^{-1/2} (\breve C_{it}^+-C_{it}^0) \dconv N(0,1), \end{equation}
where
\[ V_{it}^*=\lambdaUpperCase^{0\prime}_i \Sigma_\Lambda^{-1}\Gamma_t^* \Sigma_\Lambda^{-1}\lambdaUpperCase_i^0, \quad
 W_{it}= \Fop_t(\Sigma_F^{-1} \Phi_i \Sigma_F^{-1}) F_t^0 \]
and $\Gamma_t^*$ is defined in (\ref{Gammat*}). Note that we assume the two terms in the representation are asymptotically independent so that
the variance is the sum of the two variances.

A useful special case occurs, if for a given entry $(i,t)$,  the corresponding row and column are observable
(i.e., if  ($X_{i1},X_{i2},...,X_{iT})$ and
$(X_{1t},X_{2t},...,X_{Nt})$ are observable), then
\[  \Big(\frac 1 N V_{it}+ \frac 1 T W_{it}\Big)^{-1/2} (\breve C_{it}^+-C_{it}^0) \dconv N(0,1), \]
where $ V_{it}=\lambdaUpperCase^{0\prime}_i \Sigma_\Lambda^{-1}\Gamma_t \Sigma_\Lambda^{-1}\lambdaUpperCase_i^0$. This follows from
$N_{o_t}=N$,  $B_t^i=I_r$,  $\Gamma_t^*=\Gamma_t$, and $T_{0_i}=T$.





We remark that using the convergence rate for $\tilde C_{it}^+-C_{it}^+$, we can also show, under the assumption that $N_{o_t}\ge N_o$, and $T_\oi\ge T_o$,
the Frobenius norm of the matrix $\tilde C^+-C^0$ is bounded by
\begin{equation}\label{breveC-C-norm} \frac {\|\breve C^+-C^0\|} {\sqrt{N T }} = O_p(\frac 1 {\sqrt{N}})+  O_p(\frac 1 {\sqrt{T}} )+\sqrt{(1-p_N)(1-p_T)}\Big[ O_p( \frac 1 {\sqrt{N p_N}}) +
O_p( \frac 1 {\sqrt{T p_T}})\Big]   \end{equation}
where $p_N=N_0/N$ and $p_T=T_0/T$.

\paragraph{Proof of Lemma \ref{lemma-S-S}: }  Omitting the subscript $X$, we write $\Sigma_{ij}$ for $\Sigma_{X,ij}$, and similarly for the estimated
counterpart. For $i\ne j$,
\begin{align*}
\hat \Sigma_{ij}-\Sigma_{ij} & =  \tilde \Lambda_i'\tilde\Sigma_F \tilde \Lambda_j-\Lambda_i^{0'}\Sigma_F \Lambda_j^0 \\
& =(\tilde \Lambda_i-G\Lambda_i^0)'\tilde \Sigma_F \tilde \Lambda_j +\Lambda_i^{0'} G' \tilde \Sigma_F G G^{-1}\tilde \Lambda_j-\Lambda_i^{0'}\Sigma_F \Lambda_j^0 \\
& =(\tilde \Lambda_i-G\Lambda_i^0)'\tilde \Sigma_F \tilde \Lambda_j +\Lambda_i^{0'} (G '\tilde \Sigma_F G-\Sigma_F) G^{-1}\tilde \Lambda_j
+ \Lambda_i^{0'}\Sigma_F (G^{-1}\tilde \Lambda_j -\Lambda_j^0),
\end{align*}
where $G$ is either $H_{tall}^{-1}$ or $(H^+)^{-1}$ depending on the choice of $\tilde \Lambda$.
Thus
\begin{align*} \max_{ij} |\hat \Sigma_{ij}-\Sigma_{ij}|
& \le  \max_i \|\tilde \Lambda_i-G \Lambda_i^0\| \cdot \|\tilde \Sigma_F\| \max_j\|\tilde \Lambda_j\|  \\
&+(\max_i\|\Lambda_i^0\|)(\max_j\|\tilde \Lambda_j\|) \|G'\tilde \Sigma_F G-\Sigma_F\|  \|G^{-1}\|\\
&+\max_i \|\Lambda_i^0\| \|\Sigma_F G^{-1}\|
\max_j\|\tilde\Lambda_j-G\Lambda_i^0\|.  \end{align*}
By Assumption A, $\max_i\|\Lambda_i^0\|$ is bounded. Under exponential tails for the idiosyncratic errors $e_{it}$ (e.g, Fan et al, 2011), then it can be shown that
\[ \max_i \|\tilde \Lambda_i-G \Lambda_i^0\|  \le  \frac { \log (\max\{T,N\})  } {\sqrt {\min \{N_o,T_o\} }  } O_p(1). \]
By adding and subtracting terms, we have $\max_j\|\tilde \Lambda_j\|=O_p(1)$.
Note $\Sigma_F$ is a fixed dimensional matrix, we have  $\|G'\tilde \Sigma_F G-\Sigma_F\|=o_p(1)$.
In summary, for $i\ne j$,
\[ \max_{ij} |\hat \Sigma_{ij}-\Sigma_{ij}| =o_p(1). \]
For $i=j$, we need to show further that the idiosyncratic variance estimator is uniformly consistent over $i$,\begin{equation} \label{di-di}  \max_i |\hat {\Psi}_{e,ii}^2-{\Psi}_{e,ii}^2 | =o_p(1),  \end{equation}
where $\Psi_{e,ii}^2=var(e_{it})$, and $\hat {\Psi}_{e,ii}^2$ is its estimator.
From $\tilde e_{it}=e_{it}+ \tilde C_{it}-C_{it}$, we have
\[  \max_i |\hat {\Psi}_{e,ii}^2-{\Psi}_{e,ii}^2 | \le \max_i | \frac 1 {T_{o_i}} \sum_s  e_{is }^2 -{\Psi}_{ii}^2|+ 2 \max_i \frac 1 {T_{o_i}} \sum_s  |\tilde C_{is }-C_{is}| |e_{is}|+
\max_i \frac 1 {T_{o_i}} \sum_s  (\tilde C_{is }-C_{is})^2  \]
where the sum is based on $T_{o_i}$ number of entries.
Using the convergence of $\tilde  C_{it}$, and the exponential tails of $e_{it}$, we can show that each of the right  hand side term
is $o_p(1)$, implying  (\ref{di-di}).  Lemma \ref{lemma-S-S} is obtained by  combining results.


\end{document}